\documentclass[aps,pra,twocolumn,superscriptaddress,showpacs,nofootinbibfloatfix,amsmath,amsfonts,amssymb]{revtex4-2}%
\usepackage{amsmath,amsfonts,amssymb,color}
\usepackage{amsthm}
\usepackage{leftidx}
\usepackage{graphicx}
\usepackage{xcolor}
\usepackage{dcolumn}
\usepackage{bm}
\usepackage{epstopdf}
\usepackage{epsfig}
\usepackage{environ}
\usepackage{pdfcomment}

\usepackage{multirow}
\usepackage{setspace}
\usepackage{color}

\usepackage{float}
\usepackage[T1]{fontenc}
\usepackage[latin9]{inputenc}
\usepackage{setspace}
\usepackage{esint}

\begin{document}


\title{Entanglement phase transitions in non-Hermitian Floquet systems}

\author{Longwen Zhou}
\email{zhoulw13@u.nus.edu}
\affiliation{%
	College of Physics and Optoelectronic Engineering, Ocean University of China, Qingdao, China 266100
}
\affiliation{%
	Key Laboratory of Optics and Optoelectronics, Qingdao, China 266100
}
\affiliation{%
	Engineering Research Center of Advanced Marine Physical Instruments and Equipment of MOE, Qingdao, China 266100
}

\date{\today}

\begin{abstract}
The competition between unitary time-evolution and quantum measurements
could induce phase transitions in the entanglement characteristics
of quantum many-body dynamics. In this work, we reveal such entanglement
transitions in the context of non-Hermitian Floquet systems. Focusing
on noninteracting fermions in a representative bipartite lattice with
balanced gain/loss and under time-periodic quenches, we uncover rich
patterns of entanglement transitions due to the interplay between
driving and non-Hermitian effects. Specially, we find that the monotonic
increase of quenched hopping amplitude could flip the system between
volume-law and area-law entangled Floquet phases, yielding alternated
entanglement transitions. Meanwhile, the raise of gain/loss strength
could trigger area-law to volume-law reentrant transitions in the
scaling behavior of steady-state entanglement entropy, which are abnormal
and highly unexpected in non-driven systems. Connections between entanglement
transitions and parity-time-reversal (PT) transitions in Floquet spectra
are further established. Our findings not only build a foundation
for exploring entanglement phase transitions in Floquet non-Hermitian
setups, but also provide efficient means to engineer and control such
transitions by driving fields.
\end{abstract}

\maketitle

\section{Introduction \label{sec:Int}}

Non-Hermitian Floquet systems have attracted great interest in recent
years (see Ref.~\cite{NHFTPRev} for a review). The interplay between periodic
drivings and non-Hermitian effects were found to generate a great
variety of phases and transitions with unique dynamical and topological
features in insulating \cite{NHFTI00,NHFTI01,NHFTI02,NHFTI03,NHFTI04,NHFTI05,NHFTI06,NHFTI07,NHFTI08,NHFTI09,NHFTI10,NHFTI11,NHFTI12,NHFTI13,NHFTI14,NHFTI15,NHFTI16},
superconducting \cite{NHFTSC01,NHFTSC02,NHFTSC03}, semimetallic
\cite{NHFSM01,NHFSM02,NHFSM03,NHFSM04,NHFSM05} and quasicrystalline
\cite{NHFQC01,NHFQC02} systems. In experiments, non-Hermitian Floquet
physics has also been explored in various steps including photonics
\cite{NHPho01,NHPho02,NHPho03,NHPho04,NHPho05,NHPho06}, acoustics
\cite{NHAco01,NHAco02}, electrical circuits \cite{NHCir01,NHCir02}
and ultracold atoms \cite{NHCdAtm01,NHCdAtm02,NHCdAtm03,NHCdAtm04}, yielding potential
applications in stabilizing topological states and controlling material
features in open systems.

Despite constant progress, the entanglement properties of non-Hermitian
Floquet matter are much less explored. As an intriguing phenomenon
of entanglement dynamics, the measurement-induced phase transitions
have garnered increasing attention since 2018 \cite{EPT01,EPT02,EPT03,EPT04,EPT05,EPTExp01,EPTExp02,EPTExp03}.
It was found that with the increase of measurement rates, the steady-state
entanglement entropy (EE) of a quantum many-body system could undergo
a volume-law to area-law transition in its scaling behavior versus
the system size, which is originated from the competition between
unitary dynamics and projective measurements \cite{EPTRev01,EPTRev02,EPTRev03}.
Recently, such entanglement phase transitions have also been explored
in the context of non-Hermitian physics \cite{EPTNH01,EPTNH02,EPTNH03,EPTNH04,EPTNH05,EPTNH06,EPTNH07,EPTDsod01,EPTDsod02,EPTDsod03},
where the development of a dissipation gap and the presence of non-Hermitian
skin effects (NHSEs) were identified as typical mechanisms of generating
these transitions. As PT-symmetry breaking in the spectrum \cite{NHFTI00}
and NHSEs \cite{NHFTI10} can both be flexibly controlled by time-periodic
driving fields, much richer patterns of entanglement phase transitions
are expected to emerge in Floquet non-Hermitian systems compared with
the static case. Furthermore, the interplay between drivings and non-Hermitian
effects may lead to unique types of entanglement transitions that
are absent in non-driven situations, which have yet to be revealed.

In this work, we address these issues by exploring entanglement phase
transitions in non-Hermitian Floquet systems. In Sec.~\ref{sec:Mod},
we introduce one ``minimal'' model of non-Hermitian Floquet system,
which corresponds to a periodically quenched Su-Schrieffer-Heeger
(SSH) \cite{SSH01,SSH02} model with balanced gain and loss on different
sublattices. We analytically obtain the Floquet spectrum of our model
and discover rich patterns of PT transitions induced by driving and
non-Hermitian effects. In Sec.~\ref{sec:EPT}, we reveal diversified
entanglement phase transitions in our model and establish the entanglement
phase diagrams by investigating the scaling law of steady-state EE
versus the system size following long-time stroboscopic evolutions.
In Sec.~\ref{sec:Sum}, we summarize our results, comment on issues
related to experiments and discuss potential future directions. The
numerical method we adopted to study entanglement dynamics of
non-Hermitian Floquet systems is sketched in Appendix \ref{sec:AppA}.
Entanglement transitions in a generalized version of our model
are briefly discussed in the Appendix \ref{sec:AppB}.

\section{Model and Floquet spectrum}\label{sec:Mod}

\begin{figure}
	\begin{centering}
		\includegraphics[scale=0.255]{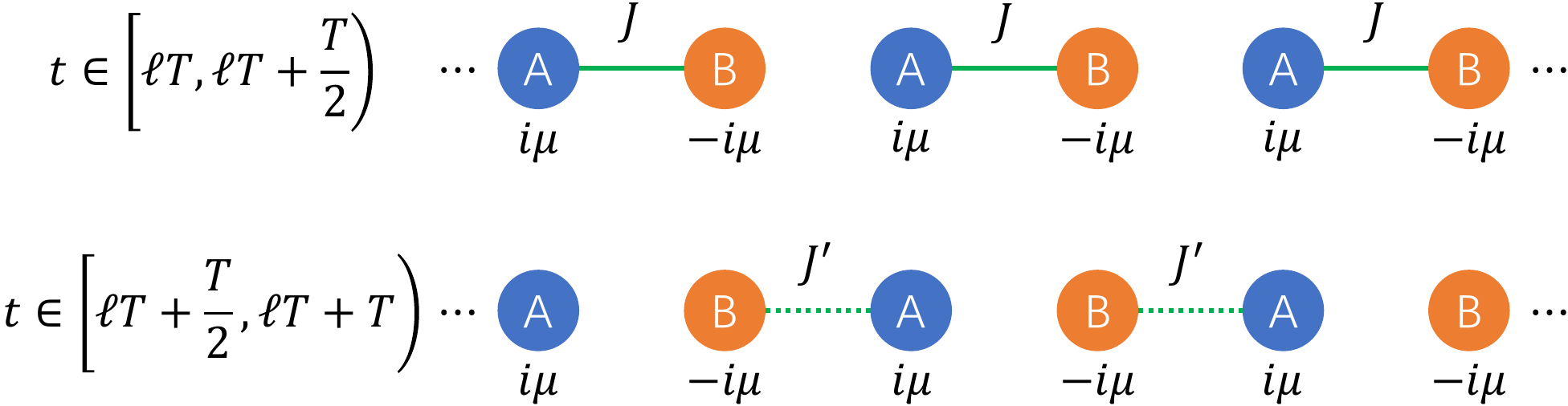}
		\par\end{centering}
	\caption{Schematic diagram of a periodically quenched non-Hermitian SSH (NHSSH)
		model. The chains in the first and second rows denote the systems
		in the first and the second half of a driving period $T$. $\ell\in\mathbb{Z}$
		and $t$ denotes time. $J$ and $J'$ are the intracell and intercell
		hopping amplitudes. Balanced gain ($i\mu$) and loss ($-i\mu$) are
		acted on the sublattices $A$ and $B$. In our calculations, we introduce
		real dimensionless parameters $J_{1}\equiv JT/(2\hbar)$, $J_{2}\equiv J'T/(2\hbar)$
		and $\gamma\equiv\mu T/(2\hbar)$ for the hopping amplitudes and gain/loss
		strength. \label{fig:Sketch}}
\end{figure}

To demonstrate the entanglement transitions in non-Hermitian Floquet
systems, we focus on a periodically quenched SSH model with balanced
gain and loss. A schematic illustration of the model is shown in 
Fig.~\ref{fig:Sketch}. 
The SSH model forms a paradigmatic setup in the study of topological insulators. It owns two sublattices $A$ and $B$ within each unit cell and possesses dimerized hopping amplitudes among adjacent lattice sites. In our driving scheme, the intracell and intercell hopping terms of the SSH model are switched on and off alternatively within each temporal modulation period. The gain and loss are time-independent and acting locally on the sublattices $A$ and $B$, respectively.
The Floquet operator of the system, which describes
its evolution over a complete driving period is given by
\begin{equation}
	\hat{U}=e^{-i\hat{H}_{2}}e^{-i\hat{H}_{1}},\label{eq:U}
\end{equation}
where
\begin{equation}
	\hat{H}_{1}=J_{1}\sum_{n}(\hat{c}_{n,A}^{\dagger}\hat{c}_{n,B}+{\rm H.c.})+i\gamma\sum_{n}(\hat{c}_{n,A}^{\dagger}\hat{c}_{n,A}-\hat{c}_{n,B}^{\dagger}\hat{c}_{n,B}),\label{eq:H1}
\end{equation}
\begin{equation}
	\hat{H}_{2}=J_{2}\sum_{n}(\hat{c}_{n,B}^{\dagger}\hat{c}_{n+1,A}+{\rm H.c.})+i\gamma\sum_{n}(\hat{c}_{n,A}^{\dagger}\hat{c}_{n,A}-\hat{c}_{n,B}^{\dagger}\hat{c}_{n,B}).\label{eq:H2}
\end{equation}
Here $\hat{c}_{n,s}^{\dagger}$ ($\hat{c}_{n,s}$) creates (annihilates)
a fermion on the sublattice $s$ ($=A,B$) in the $n$th unit cell.
Applying the Fourier transformations $\hat{c}_{n,s}=\frac{1}{\sqrt{L}}\sum_{k}e^{ikn}\hat{c}_{k,s}$
for $s=A,B$ to the system with $L$ unit cells under the PBC, we
can express $\hat{H}_{1}$ and $\hat{H}_{2}$ in the momentum space
as $\hat{H}_{1}=\sum_{k}\hat{C}_{k}^{\dagger}H_{1}(k)\hat{C}_{k}$
and $\hat{H}_{2}=\sum_{k}\hat{C}_{k}^{\dagger}H_{2}(k)\hat{C}_{k}$,
where $\hat{C}_{k}^{\dagger}\equiv(\hat{c}_{k,A}^{\dagger},\hat{c}_{k,B}^{\dagger})$,
\begin{equation}
	H_{1}(k)=J_{1}\sigma_{x}+i\gamma\sigma_{z},\label{eq:H1k}
\end{equation}
\begin{equation}
	H_{2}(k)=J_{2}\cos k\sigma_{x}+J_{2}\sin k\sigma_{y}+i\gamma\sigma_{z}.\label{eq:H2k}
\end{equation}
$\sigma_{x,y,z}$ are Pauli matrices in their usual representations.
$k\in[-\pi,\pi)$ denotes the quasimomentum. The associated Floquet
operator then reads
\begin{equation}
	\hat{U}=\sum_{k}\hat{C}_{k}^{\dagger}U(k)\hat{C}_{k},\qquad U(k)=e^{-iH_{2}(k)}e^{-iH_{1}(k)}.\label{eq:Uk}
\end{equation}
It is not hard to justify that the Bloch Hamiltonians $H_{1}(k)$
and $H_{2}(k)$ both possess the PT symmetry, i.e., $[{\cal PT},H_{1}(k)]=[{\cal PT},H_{2}(k)]=0$,
with the parity ${\cal P}=\sigma_{x}$ and the time-reversal ${\cal T}={\cal K}$,
where ${\cal K}$ takes the complex conjugate. Therefore, the 
Bloch-Floquet operator $U(k)$, when expressed in a symmetric time frame
as
\begin{equation}
	{\cal U}(k)=e^{-\frac{i}{2}H_{1}(k)}e^{-iH_{2}(k)}e^{-\frac{i}{2}H_{1}(k)},\label{eq:UkSTF}
\end{equation}
also possesses the PT symmetry in the sense that 
\begin{equation}
	{\cal PT}{\cal U}(k){\cal PT}={\cal U}^{-1}(k).\label{eq:PTS}
\end{equation}
The quasienergy spectrum of ${\cal U}(k)$ could thus be real in certain
parameter domains even though ${\cal U}(k)$ is not unitary. Since
$U(k)$ and ${\cal U}(k)$ are related by a similarity transformation
that does not affect the spectrum, the original system described by
$U(k)$ could also have a real quasienergy spectrum in the same parameter
regions as of ${\cal U}(k)$. Therefore, our periodically quenched
NHSSH model holds the PT symmetry and its quasienergy spectrum may
undergo real-to-complex transitions with the increase of the gain and loss
strength $\gamma$.

\begin{figure*}
	\begin{centering}
		\includegraphics[scale=0.46]{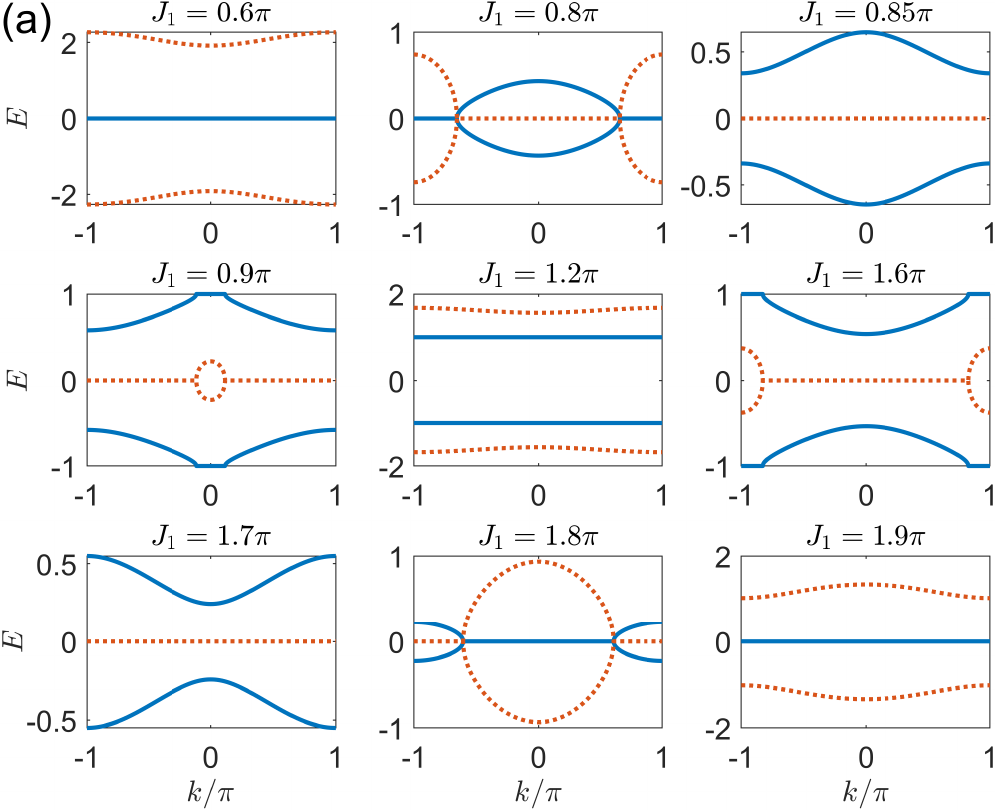}$\,\,$\includegraphics[scale=0.46]{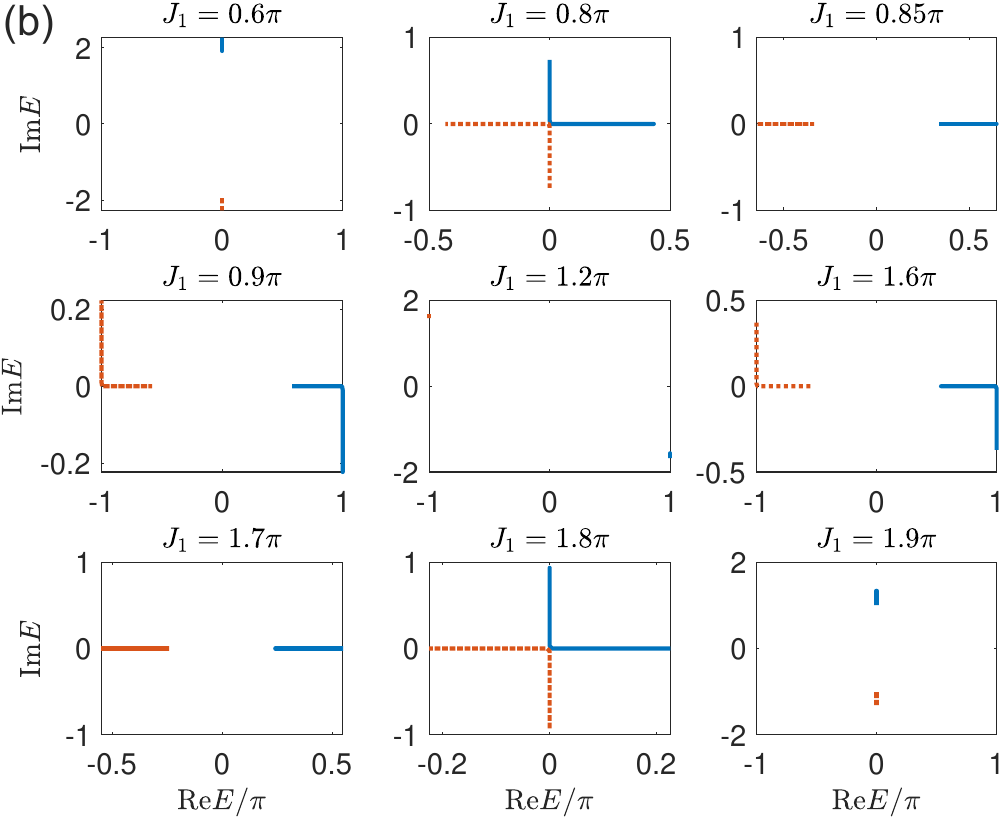}
		\par\end{centering}
	\begin{centering}
		\includegraphics[scale=0.47]{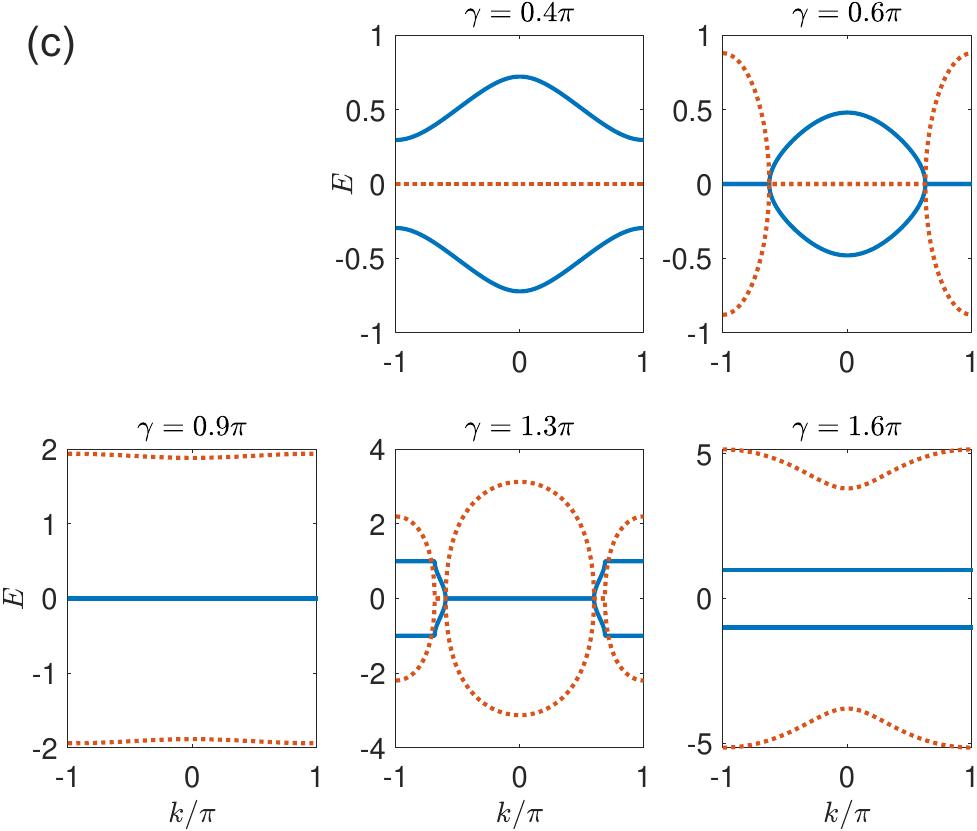}$\,\,\,$\includegraphics[scale=0.47]{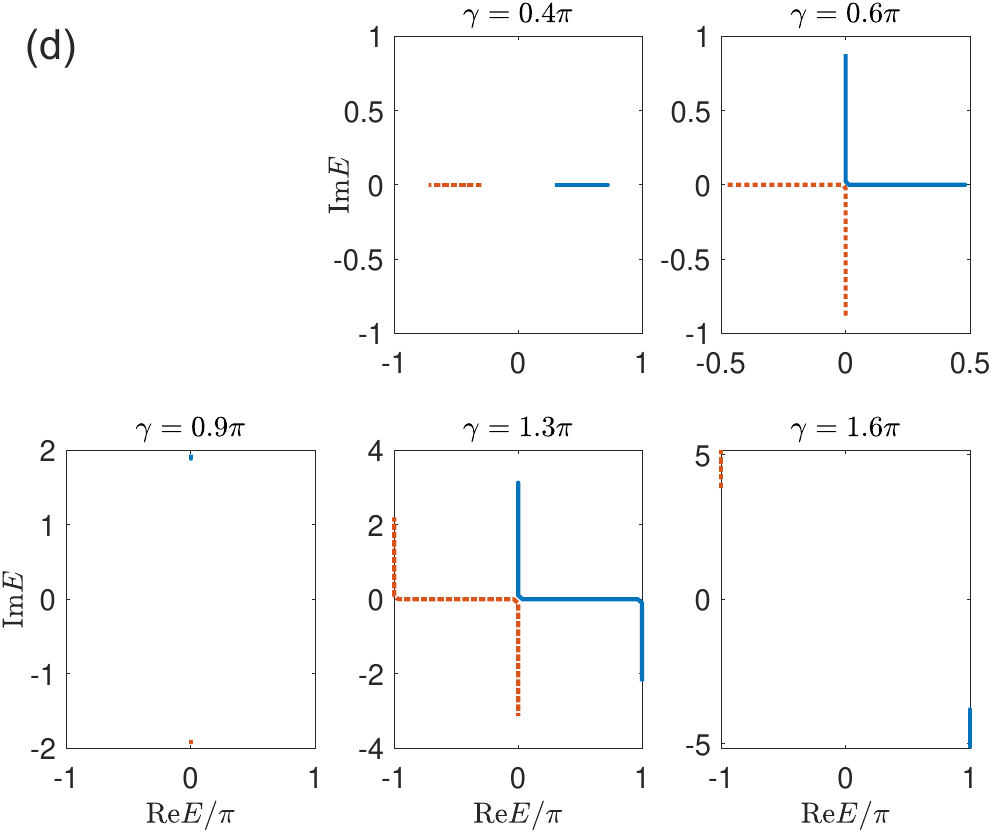}
		\par\end{centering}
	\caption{Floquet quasienergy spectrum of the periodically quenched NHSSH model
		under PBC. Other system parameters are $(J_{2},\gamma)=(0.1\pi,0.5\pi)$
		for (a), (b) and $(J_{1},J_{2})=(2.2\pi,2\pi/3)$ for (c), (d). In
		(a) and (c), the blue solid and red dotted lines denote $\pm{\rm Re}E(k)/\pi$
		and $\pm{\rm Im}E(k)$ vs the quasimomentum $k$. In (b) and (d),
		the blue solid and red dotted lines represent the two Floquet bands
		$\pm E$ on the complex quasienergy plane. \label{fig:EPBC}}
\end{figure*}

The Floquet spectrum of our system can be obtained by solving the
eigenvalue equation $U(k)|\psi\rangle=e^{-iE(k)}|\psi\rangle$. The
resulting quasienergy dispersions take the forms $E_{\pm}(k)=\pm E(k)$,
with
\begin{equation}
	E(k)=\arccos(\cos E_{1}\cos E_{2}-{\bf n}_{1}\cdot{\bf n}_{2}\sin E_{1}\sin E_{2}).\label{eq:Ek}
\end{equation}
Here the terms $E_{1}=\sqrt{J_{1}^{2}-\gamma^{2}}$,
$E_{2}=\sqrt{J_{2}^{2}-\gamma^{2}}$, ${\bf n}_{1}=(J_{1},0,i\gamma)/E_{1}$
and ${\bf n}_{2}=(J_{2}\cos k,J_{2}\sin k,i\gamma)/E_{2}$. From 
Eq.~(\ref{eq:Ek}), it is not hard to verify that $\cos[E(k)]$ is always
real. Therefore, when $|\cos[E(k)]|<1$ for all $k$, the quasienergy
dispersions $\pm E(k)$ must be purely real and the system resides
in the PT-invariant regime. When $|\cos[E(k)]|>1$ for certain $k$,
the $\pm E(k)$ must be complex and the system goes into a PT-broken
phase. A PT transition in the system is then expected to happen at
$\cos[E(k)]=\pm1$, or
\begin{equation}
	\cos E_{1}\cos E_{2}-{\bf n}_{1}\cdot{\bf n}_{2}\sin E_{1}\sin E_{2}=\pm1.\label{eq:PB}
\end{equation}
Note that these are also the conditions for the two Floquet bands
$E_{\pm}(k)$ to touch with each other at the center ($E=0$, $\cos[E(k)]=1$)
and boundary ($E=\pi$, $\cos[E(k)]=-1$) of the first quasienergy
Brillouin zone (BZ) ${\rm Re}E\in(-\pi,\pi]$, respectively.
These touching points are second-order Floquet exceptional points (FEPs). In the first BZ of $k$, they appear at the quasimomenta $\pm k_c$ that satisfy $k_{c}=\arccos\left[\frac{1}{J_{1}J_{2}}\left(\frac{\cos E_{1}\cos E_{2}-1}{\sin E_{1}\sin E_{2}}E_{1}E_{2}+\gamma^{2}\right)\right]$ with $E(k_{c})=0$ and $k_c=\arccos\left[\frac{1}{J_{1}J_{2}}\left(\frac{\cos E_{1}\cos E_{2}+1}{\sin E_{1}\sin E_{2}}E_{1}E_{2}+\gamma^{2}\right)\right]$ with $E(k_{c})=\pi$, respectively. Note that the FEPs emerging at the boundary of Floquet BZ ($E=\pi$) are unique to non-Hermitian Floquet systems. Similar to exceptional points of static Hamiltonians \cite{EP01}, FEPs could also lead to the breakdown of adiabatic predictions in driven systems.

In Fig.~\ref{fig:EPBC}, we present typical cases of the Floquet spectrum
$E_{\pm}(k)$ {[}Eq.~(\ref{eq:Ek}){]} for our periodically quenched
NHSSH model under PBC. We find that with the change of the hopping
or gain and loss parameter, the line quasienergy gap between the two Floquet
bands could close at either $E=0$ or $E=\pi$, which is followed
by the change of spectral compositions (real, purely complex or partially
real). We thus expect to have both PT breaking and restoring transitions
in the system, which are clearly illustrated by the panels at
different $J_{1}$ in Figs.~\ref{fig:EPBC}(a) and \ref{fig:EPBC}(b). Moreover,
we observe that with the increase of $\gamma$, the Floquet spectrum
does not change monotonically from real to purely complex. It could
instead enter an intermediate phase with both real and complex quasienergies,
as illustrated by the case with $\gamma=1.3$ in Figs.~\ref{fig:EPBC}(c)
and \ref{fig:EPBC}(d). These rich spectral patterns, as identifiable
from Fig.~\ref{fig:EPBC}, clearly distinguish our Floquet model from
its static non-Hermitian counterpart \cite{EPTNH06}. They also underline the alternated and reentrant entanglement transitions we are going to reveal in the next section.

To further characterize the composition of Floquet spectrum and discriminate
between the PT-invariant and PT-broken phases, we introduce the ratio
of real quasienergies of $U(k)$, which is defined as
\begin{equation}
	R=\int_{-\pi}^{\pi}\frac{dk}{2\pi}\Theta(1-|\cos[E(k)]|).\label{eq:R}
\end{equation}
Here $\Theta(x)$ is the step function, which is equal to $1$ ($0$) if $x>0$ ($x<0$). It is clear that we have
$R=1$ ($R=0$) if all the quasienergies of $U(k)$ are real (complex).
If $R\in(0,1)$, real and complex quasienergies coexist in the 
Floquet spectrum of $U(k)$. A PT-breaking transition then happens when the
value of $R$ starts to decrease from one.

\begin{figure}
	\begin{centering}
		\includegraphics[scale=0.48]{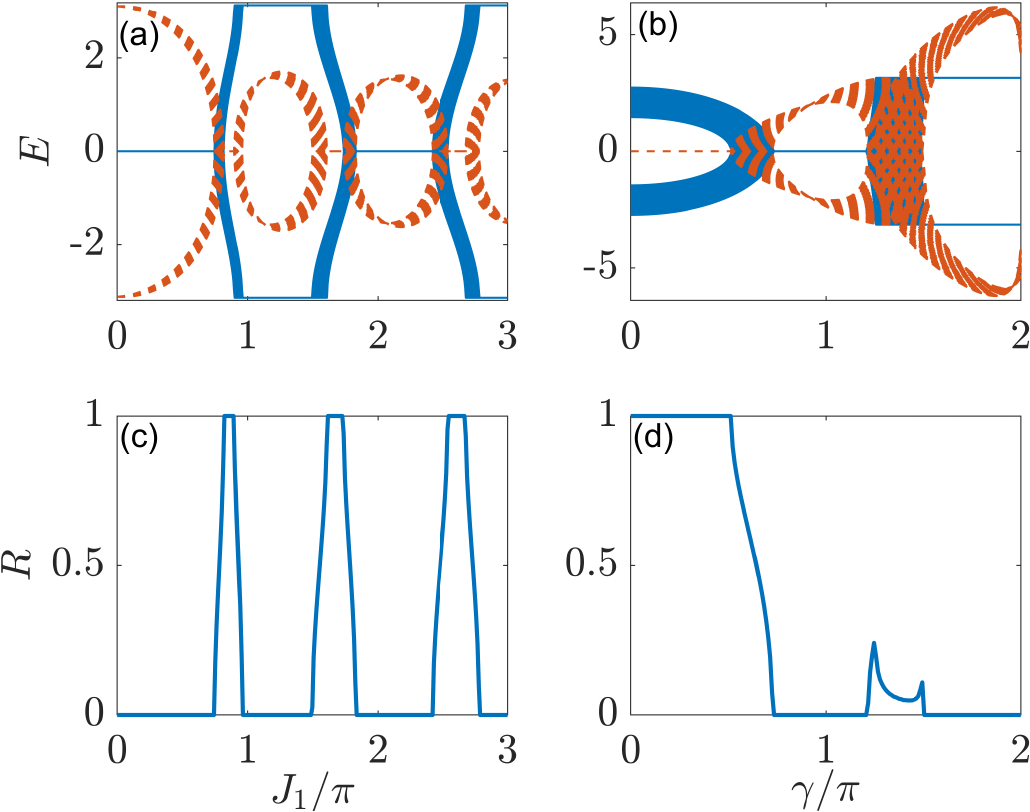}
		\par\end{centering}
	\caption{Floquet spectrum $E$ {[}(a), (b){]} and ratios of real quasienergies
		$R$ {[}(c), (d){]} vs the hopping amplitude $J_{1}$ and gain/loss
		parameter $\gamma$ under the PBC. Other system parameters are $(J_{2},\gamma)=(0.1\pi,0.5\pi)$
		for (a), (c) and $(J_{1},J_{2})=(2.2\pi,2\pi/3)$ for (b), (d). The
		solid and dashed lines in (a) and (b) denote the real and imaginary
		parts of quasienergy. \label{fig:E-R-PBC}}
\end{figure}
In Fig.~\ref{fig:E-R-PBC}, we present the quasienergy spectrum 
{[}Eq.~(\ref{eq:Ek}){]} and the real-quasienergy ratio {[}Eq.~(\ref{eq:R}){]}
versus the hopping amplitude $J_{1}$ and the gain/loss strength $\gamma$
separately for our periodically quenched NHSSH model. A series of
alternated PT-breaking ($R=1\rightarrow R<1$) and PT-restoring ($R<1\rightarrow R=1$)
transitions can be observed with the change of $J_{1}$. These
transitions are accompanied by band touchings at the quasienergy zero
or $\pi$. They are further mediated by critical phases with coexisting
real and complex quasienergies ($0<R<1$) in the Floquet spectrum.
Meanwhile, we notice that with the raise of $\gamma$ from zero, the
system could first undergo a PT-breaking transition and its Floquet
spectrum changes gradually from partially real to purely complex.
However, real quasienergies could reappear in a region with larger
$\gamma$, which is rarely achievable with the raise of gain and loss
strengths in static non-Hermitian systems. These reentrant PT transitions
and gain/loss-induced real quasienergies are both originated from
the interplay between drivings and non-Hermitian effects. Their notable
influences on entanglement phase transitions in our non-Hermitian
Floquet system will be revealed in Sec.~\ref{sec:EPT}.

\begin{figure}
	\begin{centering}
		\includegraphics[scale=0.236]{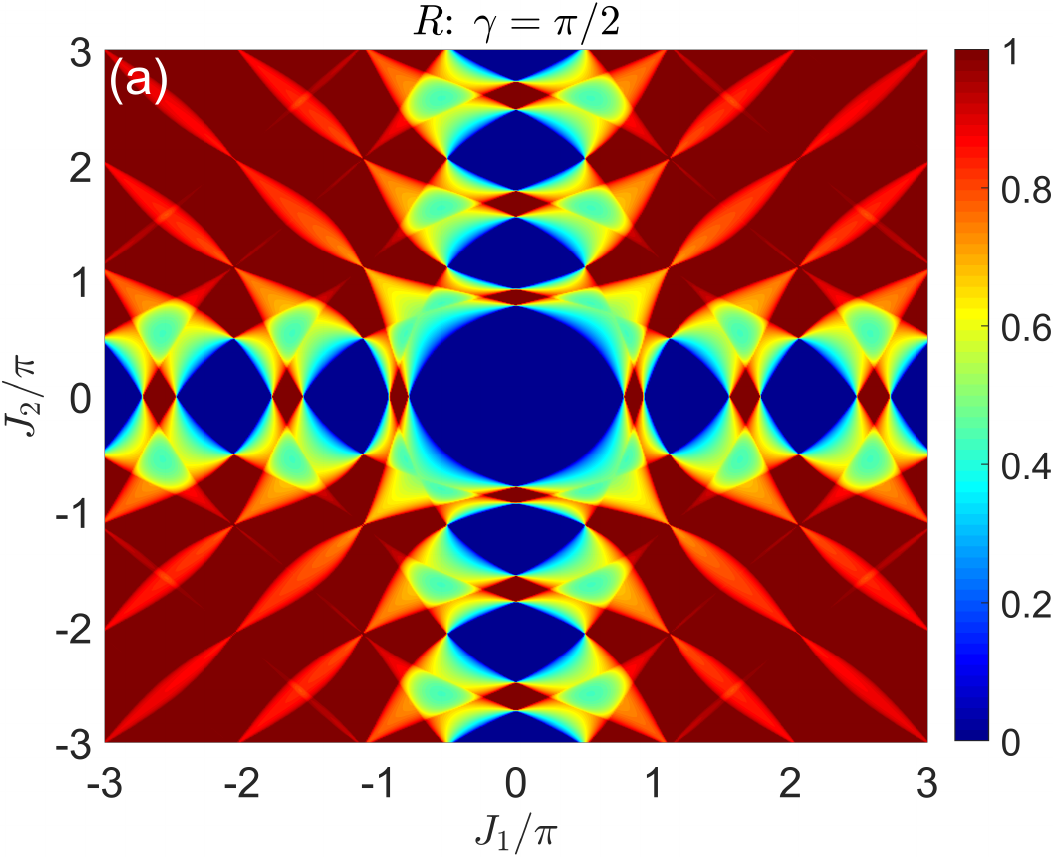}\includegraphics[scale=0.236]{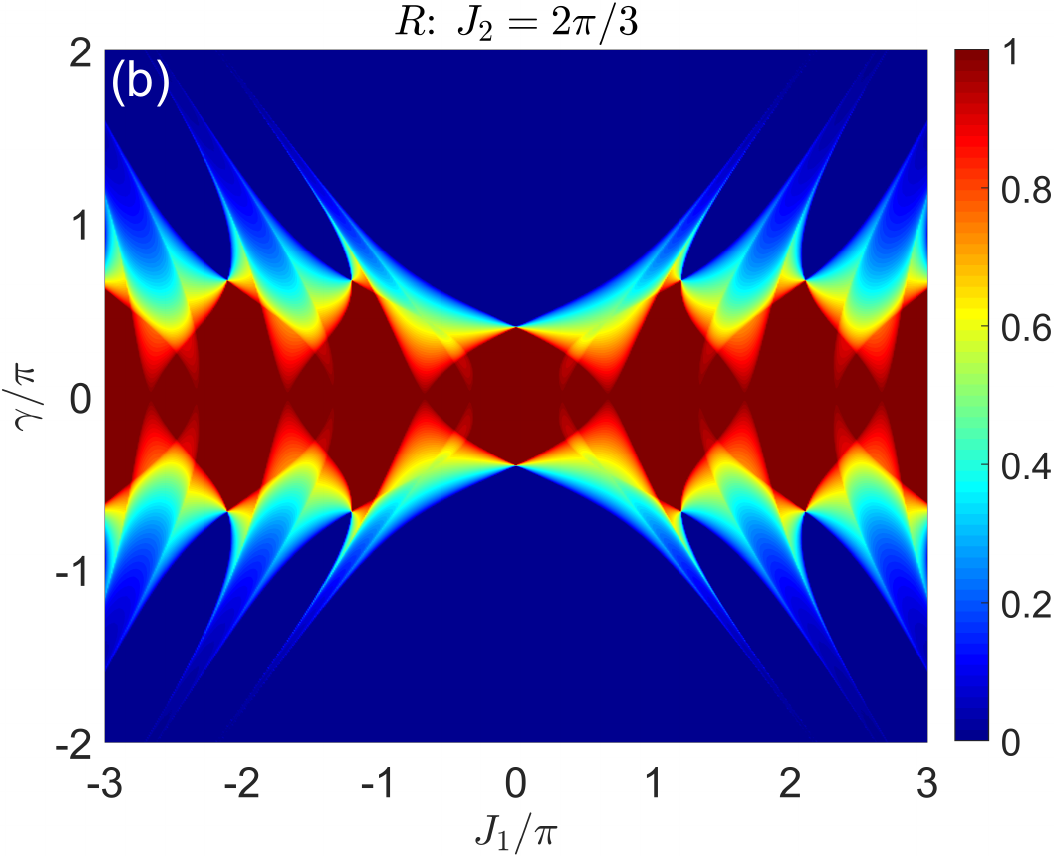}
		\par\end{centering}
	\caption{Ratios of real quasienergies $R$ {[}Eq.~(\ref{eq:R}){]} versus (a)
		$(J_{1},J_{2})$ at $\gamma=\pi/2$ and (b) $(J_{1},\gamma)$ at $J_{2}=2\pi/3$
		for the periodically quenched NHSSH model {[}Eq.~(\ref{eq:U}){]}
		under PBC. Different colors represent different values of $R$, as
		can be read out from the color bars. \label{fig:PTPhsDiag}}
\end{figure}
In Fig.~\ref{fig:PTPhsDiag}, we show the values of $R$ versus $(J_{1},J_{2})$
and $(J_{1},\gamma)$ as two typical cases of PT phase diagrams under
the PBC. In both Figs.~\ref{fig:PTPhsDiag}(a) and \ref{fig:PTPhsDiag}(b),
we observe rich patterns of PT-invariant (in dark red), PT-broken
(in dark blue) and intermediate ($0<R<1$) phases with different compositions
of the Floquet spectrum. Moreover, the change of each system parameter
could induce a series of alternated and reentrant PT transitions in
the spectrum, which is usually unavailable in static non-Hermitian
systems. Specially, we find that with the increase of gain and loss strength
$\gamma$, the spectrum could first change gradually from real ($R=1$)
to purely complex ($R=0$), and then going back to a mixed case ($0<R<1$)
with coexisting real and complex quasienergies {[}Fig.~\ref{fig:PTPhsDiag}(b){]}.
This is again unexpected in static non-Hermitian systems, where a
stronger gain and loss usually prefer a larger proportion of complex
eigenenergy in the spectrum. These observations imply that time-periodic
driving fields could not only induce rich PT phases and transitions,
but also provide a mechanism to stabilize PT-symmetric non-Hermitian
systems in stronger dissipation regions.

In the next section, we will characterize the entanglement nature
of Floquet phases with different spectral properties in our periodically
quenched NHSSH model. The rich and alternated spectrum transitions found
here will be further related to reentrant entanglement transitions
in our system.

\section{Entanglement phase transitions \label{sec:EPT}}

In static non-Hermitian systems, it has been identified that the opening
of a dissipation gap along the imaginary-energy axis could lead to
a volume-law to area-law phase transition in the EE of free fermions \cite{EPTNH06}. The non-Hermitian skin effect
constitutes another mechanism of generating such entanglement phase
transitions \cite{EPTNH04}. The presence of random or quasiperiodic
disorder may further collaborate with non-Hermitian effects to generate
anomalous log-law to area-law entanglement transitions \cite{EPTDsod02}.
Beyond these static situations, we will now demonstrate how entanglement
phase transitions could be induced and controlled by time-periodic
drivings in our non-Hermitian Floquet system.

We first outline the methodology of obtaining the EE and its stroboscopic
dynamics for noninteracting fermions in a Floquet non-Hermitian lattice.
Let us consider a system initialized in the state $|\Psi_{0}\rangle$
at time $t=0$ and evolved by the Floquet operator $\hat{U}$ with the
driving period $T$. The normalized state of the system after a number
of $\ell$ $(\in\mathbb{N})$ driving periods is given by
\begin{equation}
	|\Psi(\ell T)\rangle=\frac{\hat{U}^{\ell}|\Psi_{0}\rangle}{\sqrt{\langle\Psi_{0}|\hat{U}^{\dagger\ell}\hat{U}^{\ell}|\Psi_{0}\rangle}}.\label{eq:Psit}
\end{equation}
Note in passing that for a non-Hermitian Floquet system, we usually
have $\hat{U}^{\dagger}\hat{U}\neq1$, and the resulting stroboscopic
dynamics is not unitary.
Physically, the re-normalization process in Eq.~(\ref{eq:Psit}) can be related to open-system dynamics dependent on measurement outcomes, so that the quantum evolution corresponds to the null-jump process \cite{EPTNH04,EPTNH06,NHDynamics01,NHDynamics02,NHDynamics03,NHDynamics04}. Over each driving period, the state evolves following the Schr\"odinger equation with an effective non-Hermitian Hamiltonian ${\hat H}=i{\rm ln}{\hat U}$. This is followed by a normalization of the wave function without going through quantum jumps (post-selection).

In the absence of gain and loss ($\gamma=0$), the Hamiltonians ${\hat H}_1$ and ${\hat H}_2$ of our system in the two halves of its evolution period are both Hermitian, and the resulting Floquet dynamics as described by ${\hat U}$ [Eq.~(\ref{eq:U})] is strictly unitary. In this case, the denominator of Eq.~(\ref{eq:Psit}) is always one for a normalized $|\Psi_0\rangle$, and there are no issues concerning the exponential growth of amplitude of the evolving state ${\hat U}^\ell |\Psi_0\rangle$. With finite gain and loss ($\gamma\neq0$), the Floquet spectrum of ${\hat U}$ may become fully complex or partially complex with second-order FEPs (see Fig.~\ref{fig:EPBC}) under PBC. In this case, the norm of the evolved state ${\hat U}^\ell |\Psi_0\rangle$ may become larger than one and even grow exponentially before the re-normalization operated in Eq.~(\ref{eq:Psit}). Possible issues related to this exponential growth of state's amplitude can be resolved by adding to our system a time-independent global loss term in the form of $-i\Gamma\sum_n({\hat c}^\dagger_{n,A}{\hat c}_{n,A}+{\hat c}^\dagger_{n,B}{\hat c}_{n,B})$, where the loss rate $\Gamma>0$ can be chosen to be large enough. In this case, the norm of the evolving state ${\hat U}^\ell |\Psi_0\rangle$ can only decay over time. Meanwhile, the dynamics described by our Eq.~(\ref{eq:Psit}) is unaffected, as the global loss terms in the numerator and denominator would be canceled.

In our numerical calculations, we take the
PBC and choose the initial state as 
\begin{equation}
	|\Psi_{0}\rangle=\prod_{n=1}^{L}\hat{c}_{n,B}^{\dagger}|\emptyset\rangle.\label{eq:Psi0}
\end{equation}
Here $L$ is the total number of unit cells in the lattice and $|\emptyset\rangle$
denotes the vacuum state. The state $|\Psi_{0}\rangle$ in Eq.~(\ref{eq:Psi0})
thus describes a charge density wave at half-filling, with each sublattice
$B$ being populated by a single fermion. 
This initial state is not an eigenstate of the Floquet operator ${\hat U}$ [Eq.~(\ref{eq:U})] or the noncommutative Hamiltonians ${\hat H}_{1,2}$ [Eqs.~(\ref{eq:H1})--(\ref{eq:H2})]. The system's hopping parameters will experience sudden changes in the middle of each driving period, and there are no slowly varying parameters. Therefore, our system will undergo far-from-equilibrium dynamics and reaching some steady states after being evolved stroboscopically over multiple driving periods. Such a process is nonadiabatic and thus could not be captured by adiabatic predictions.
Other types of pure and nonequilibrium initial states
yield consistent results concerning the entanglement transitions that
will be studied below.

At any stroboscopic time $t=\ell T$, the matrix
elements of single-particle correlator $C(\ell T)$ in the lattice
representation can now be expressed as
\begin{equation}
	C_{ms,m's'}(\ell T)=\langle\Psi(\ell T)|\hat{c}_{m,s}^{\dagger}\hat{c}_{m',s'}|\Psi(\ell T)\rangle,\label{eq:CM}
\end{equation}
where $m,m'=1,...,L$ and $s,s'=A,B$ denote the unit cell and sublattice
indices, respectively. Restricting the indices $m,m'$ of $C(\ell T)$
to a subsystem X with $l$ unit cells gives us a $2l\times2l$ block
of $C(\ell T)$. The eigenvalues of this block constitute the correlation-matrix
spectrum $\{\zeta_{j}(\ell T)|j=1,...2l\}$ of the subsystem X. Without
interactions, the $|\Psi(t)\rangle$ is a Gaussian state and the bipartite
EE can be obtained from the spectrum of correlation matrix \cite{PeschelRev2009}.
That is, at any given stroboscopic time $t=\ell T$, we can find the
EE between the subsystem X and remaining part Y of the whole system
as
\begin{equation}
	S(t)=-\sum_{j=1}^{2l}[\zeta_{j}\ln\zeta_{j}+(1-\zeta_{j})\ln(1-\zeta_{j})].\label{eq:EEt}
\end{equation}
Here we have suppressed the time-dependence in $\zeta_{j}$ for brevity.
The $S(t)$ thus defined corresponds to the bipartite EE $S(t)=-{\rm Tr}[\rho_{{\rm X}}(t)\ln\rho_{{\rm X}}(t)]$,
where the reduced density matrix $\rho_{{\rm X}}(t)$ of subsystem
X can be obtained by tracing out all the degrees of freedom belonging
to the remaining subsystem Y with $2(L-l)$ sites, i.e., $\rho_{{\rm X}}(t)={\rm Tr}_{{\rm Y}}[|\Psi(t)\rangle\langle\Psi(t)|]$.
The numerical recipe of computing the single-particle correlation matrix and EE for a non-Hermitian Floquet
system are summarized in Appendix \ref{sec:AppA}. We emphasize that our approach only uses the right vector of a single wave function, instead of biorthogonal density matrices constructed from both right and left eigenstates.

\begin{figure}
	\begin{centering}
		\includegraphics[scale=0.5]{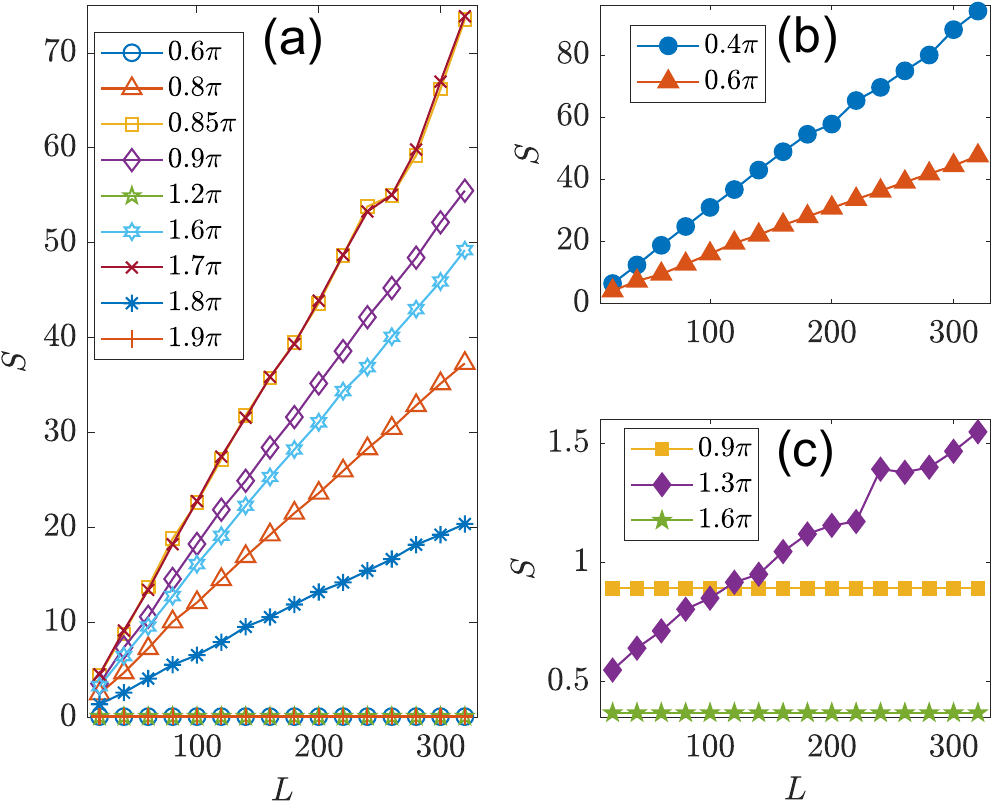}
		\par\end{centering}
	\caption{Steady-state EE $S(L,l)$ versus the system size $L$ under PBC for
		a equal bipartition $l=L/2$ and at half-filling. In (a), the legend
		gives the value of $J_{1}$ for each curve, with other system parameters
		given by $(J_{2},\gamma)=(0.1\pi,0.5\pi)$ {[}same as those taken
		in Figs.~\ref{fig:EPBC}(a) and \ref{fig:EPBC}(b){]}. In (b) and
		(c), the legends show the value of $\gamma$ for each curve, with
		other system parameters given by $(J_{1},J_{2})=(2.2\pi,2\pi/3)$
		{[}same as those used in Figs.~\ref{fig:EPBC}(c) and \ref{fig:EPBC}(d){]}.
		\label{fig:EEvsL}}
\end{figure}
We first investigate the scaling behaviors of steady-state EE $S(L,l)$
vs the system size $L$ for $l=L/2$ (equal bipartition), at half-filling
and under PBC. For a given $L$ and $l$, $S(L,l)$ is obtained by
averaging the stroboscopic EE $S(t)$ {[}Eq.~(\ref{eq:EEt}){]} over
a late-time domain $t\in[\ell'T,\ell T]$ with $1\ll\ell'<\ell$,
where we take $\ell'=800$ and $\ell=1000$ throughout our numerical
calculations. In Fig.~\ref{fig:EEvsL}, we observe two drastically
distinct scaling behaviors in $S(L,L/2)$ at different strengths of
hopping $J_{1}$ {[}Fig.~\ref{fig:EEvsL}(a){]} and gain/loss $\gamma$
{[}Figs.~\ref{fig:EEvsL}(b)--(c){]}. Referring to Fig.~\ref{fig:EPBC},
we realize that whenever the Floquet spectrum of our periodically
quenched NHSSH model forms a dissipation gap along the imaginary-quasienergy
axis, the steady-state EE becomes independent of the system size $L$
in Fig.~\ref{fig:EEvsL}, such that $S(L,L/2)\sim L^{0}$ and area-law
scalings are observed in associated cases. Instead, in the cases when
imaginary quasienergy gaps vanish in Fig.~\ref{fig:EPBC}, the steady-state
EE becomes proportional to the system size $L$ in Fig.~\ref{fig:EEvsL},
such that volume-law entangled phases with $S(L,L/2)\sim L$ are reached. 

Note in passing that within the volume-law entangled phases, the gradients
of $S(L,L/2)$ versus $L$ reach maximal values in PT-invariant cases
with real quasienergy spectra {[}for $J_{1}=0.85\pi,1.7\pi$ in 
Fig.~\ref{fig:EEvsL}(a) and $\gamma=0.4\pi$ in Fig.~\ref{fig:EEvsL}(b){]}.
Meanwhile, volume-law scalings of the steady-state EE can be observed
in both PT-invariant and PT-broken phases, so long as there are no
dissipation gaps along the imaginary quasienergy axis. These observations
indicate that PT transitions do not have one-to-one correspondences
with entanglement transitions in non-Hermitian Floquet systems. As
another notable result, the scaling behavior of $S(L,L/2)$ changes
from area-law to volume-law when the gain and loss strength $\gamma$
raises from $0.9\pi$ to $1.3\pi$ in Fig.~\ref{fig:EEvsL}(c), which
goes beyond the situation normally expected in non-driven systems. 

\begin{figure}
	\begin{centering}
		\includegraphics[scale=0.325]{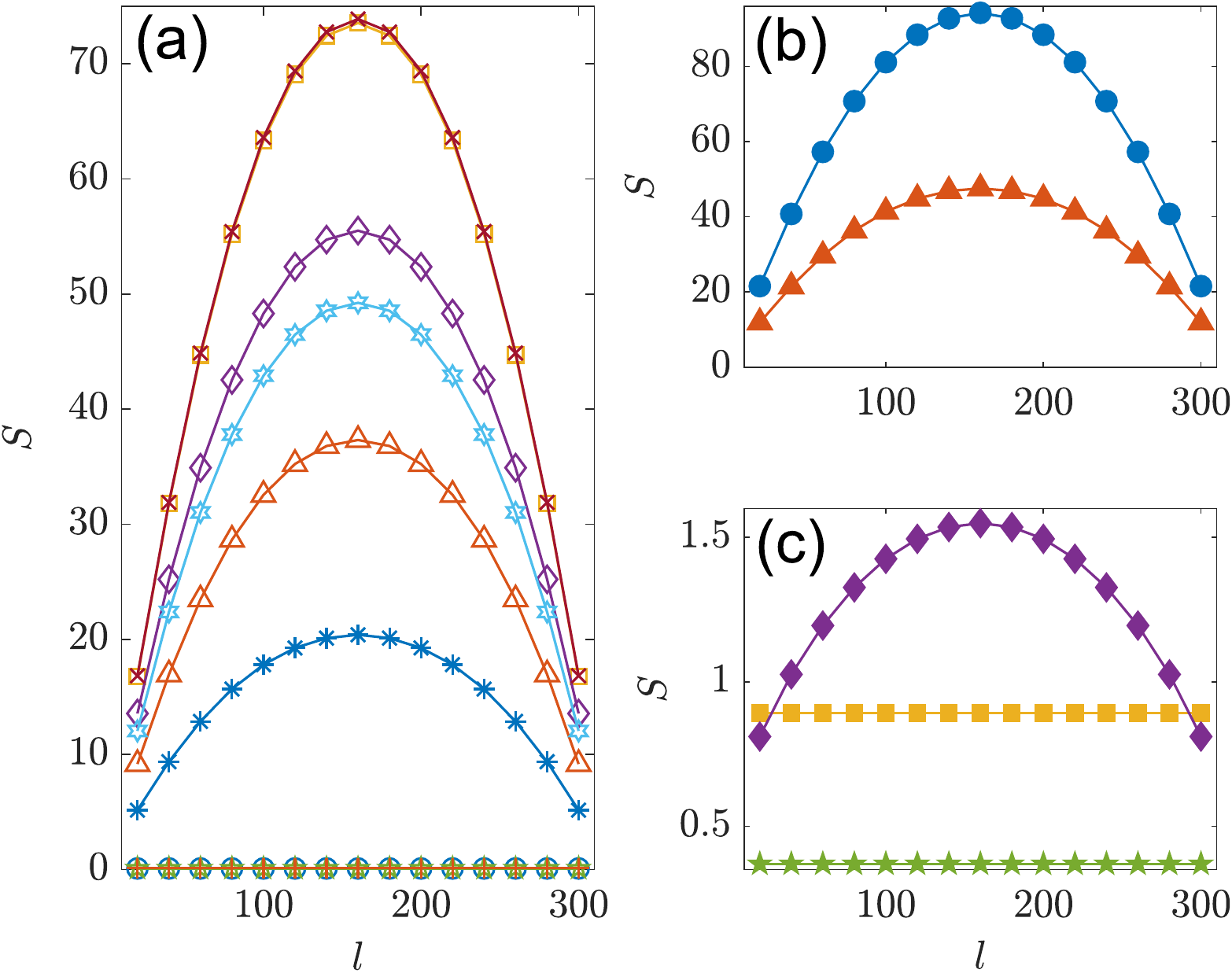}
		\par\end{centering}
	\caption{Steady-state EE $S(L,l)$ versus the subsystem size $l$ under PBC
		for a fixed total system size $L=320$ and at half-filling. (a)--(c)
		share the same legends with the corresponding panels (a)--(c) of
		Fig.~\ref{fig:EEvsL}. The curves marked by the same symbol in 
		Figs.~\ref{fig:EEvsL} and \ref{fig:EEvsLS} have the same system parameters.
		\label{fig:EEvsLS}}
\end{figure}

\begin{figure*}
	\begin{centering}
		\includegraphics[scale=0.45]{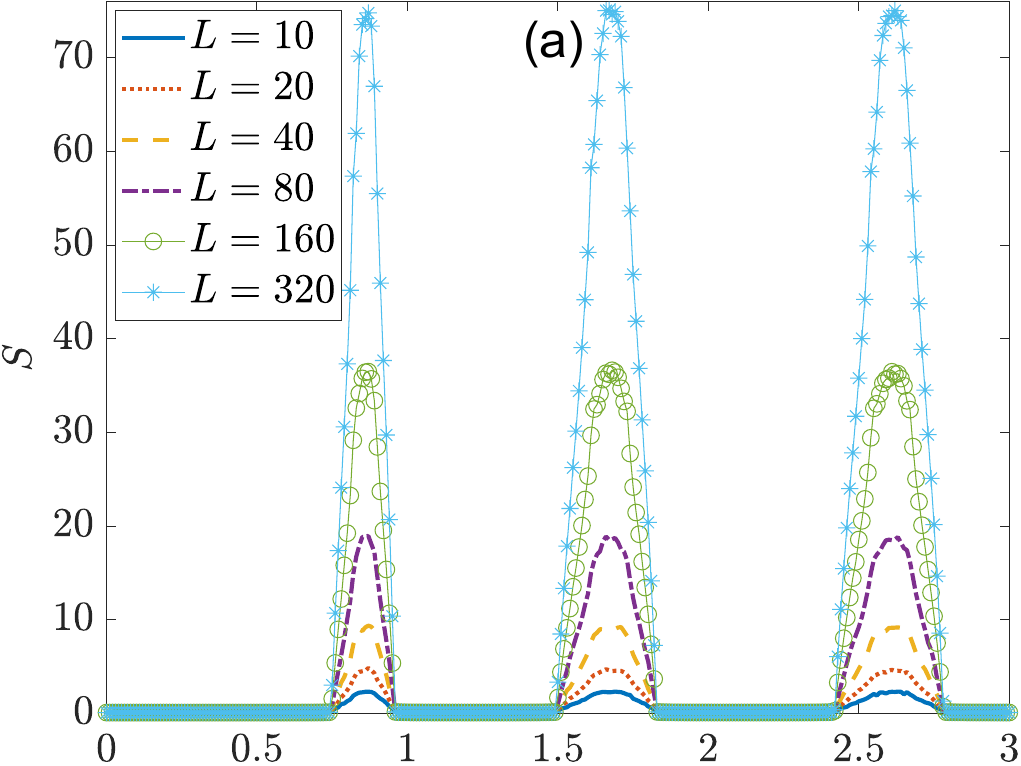}$\quad\,\,$\includegraphics[scale=0.28]{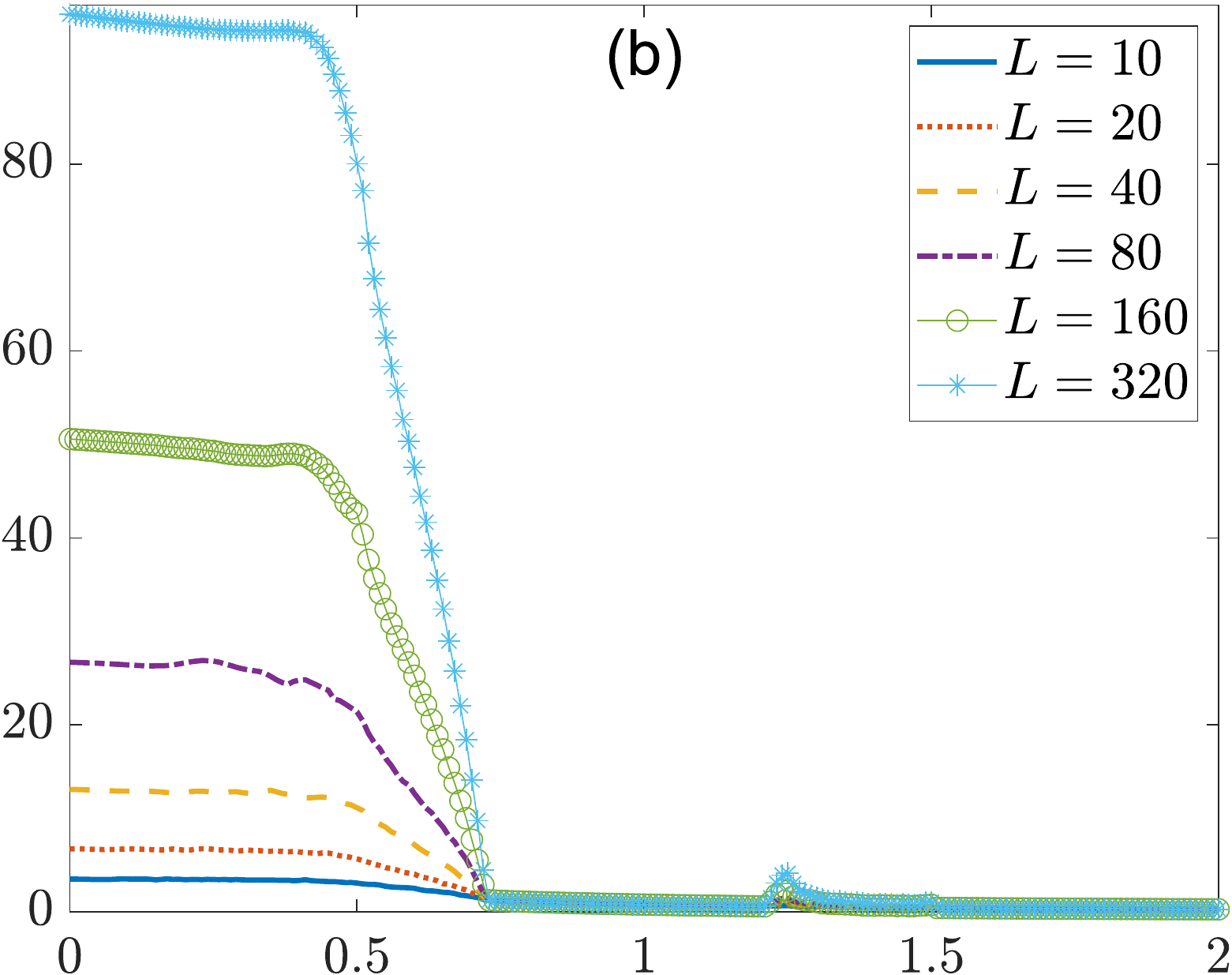}
		\par\end{centering}
	\begin{centering}
		\includegraphics[scale=0.3]{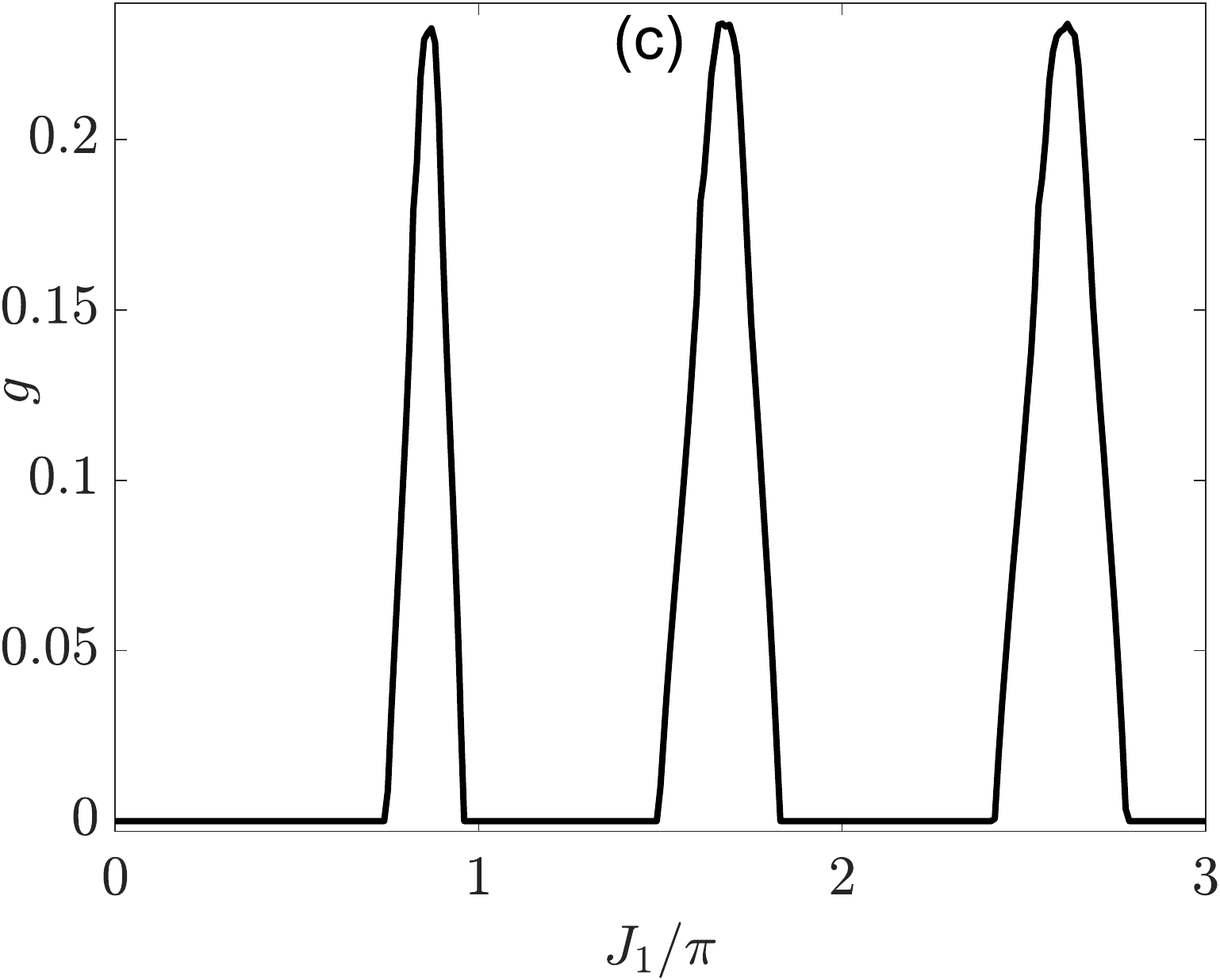}$\,\,$\includegraphics[scale=0.46]{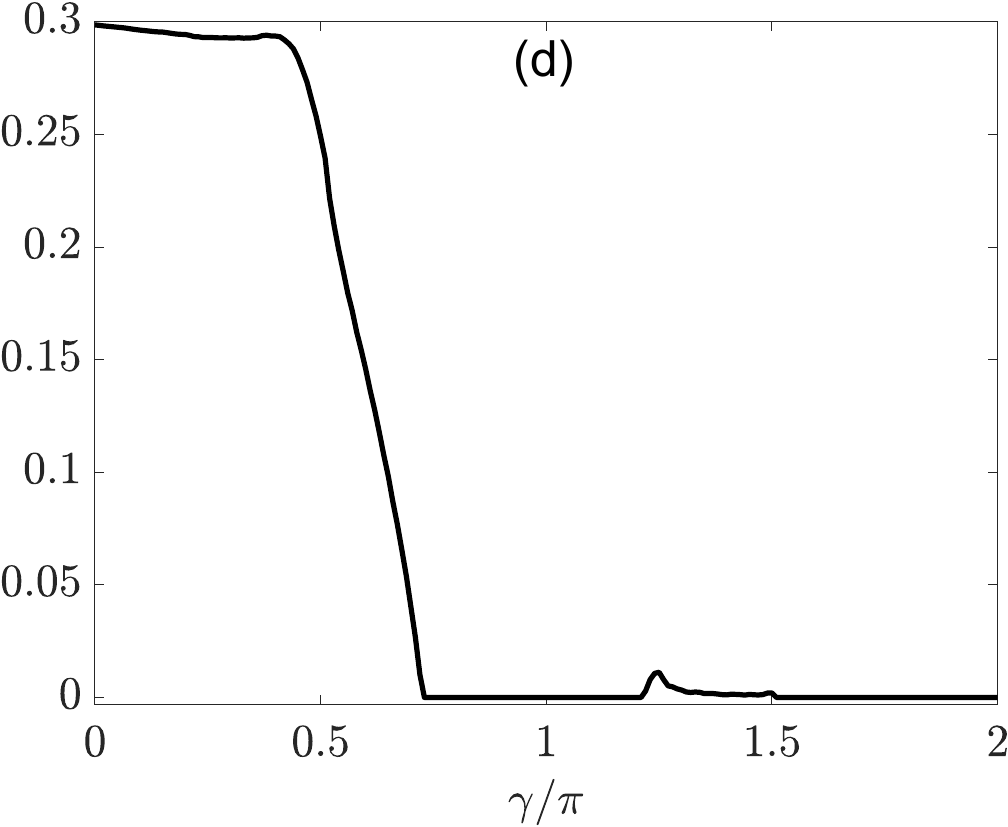}
		\par\end{centering}
	\caption{Reentrant entanglement transitions versus the hopping amplitude $J_{1}$
		{[}(a), (c){]} and gain/loss strength $\gamma$ {[}(b), (d){]}. System
		parameters are $(J_{2},\gamma)=(0.1\pi,0.5\pi)$ for (a), (c) and
		$(J_{1},J_{2})=(2.2\pi,2\pi/3)$ for (b), (d), which are the same
		as those chosen for the panels (a), (c) and (b), (d) of Fig.~\ref{fig:E-R-PBC},
		respectively. (a) and (b) show the steady-state EE $S(L,l)$, with
		$l=L/2$, versus $J_{1}$ and $\gamma$ for different lattice sizes
		$L$. (c) and (d) show the gradients $g$ extracted from the linear
		fitting $S(L,L/2)\sim gL+s_{0}$. \label{fig:EEdEE}}
\end{figure*}

To further decode entanglement phase transitions in our non-Hermitian
Floquet setting, we present in Fig.~\ref{fig:EEvsLS} the steady-state
EE $S(L,l)$ versus the subsystem size $l$ with a fixed total number
of unit cells $L=320$. The system is still at half-filling and under
PBC. In Figs.~\ref{fig:EEvsLS}(a) and \ref{fig:EEvsLS}(c), we observe
area-law scalings $S(L,l)\sim l^{0}$ in the steady-state EE for the
cases with finite dissipation gaps along imaginary quasienergy axes
of the Floquet spectra in Fig.~\ref{fig:EPBC}. When imaginary quasienergy
gaps vanish in Fig.~\ref{fig:EPBC}, we find that the $S(L,l)$ in
Figs.~\ref{fig:EEvsLS}(a)--(c) might be fitted by the function $g_{0}\sin(\pi l/L)+g_{1}\ln[\sin(\pi l/L)]+g_{2}$,
which is usually expected in volume-law entangled phases. Therefore,
the scaling behaviors of steady-state EE versus the subsystem size again
suggest two possible phases with different entanglement nature in
our periodically quenched NHSSH model, which are consistent with those
observed in Fig.~\ref{fig:EEvsL}. We also notice that the appearance
of these distinct entangling phases does not follow a monotonic sequence
with the increase of either the quenched hopping amplitude $J_{1}$
or the gain and loss strength $\gamma$.

We are now ready to demonstrate entanglement phase transitions in
our system. In Figs.~\ref{fig:EEdEE}(a) and \ref{fig:EEdEE}(b),
we present the steady-state EE $S(L,l=L/2)$ of our periodically quenched
NHSSH model versus $J_{1}$ and $\gamma$, respectively, for several
different system sizes $L$ under the PBC and at half-filling. Two
clearly distinct regions are observed in both figures. Referring to
the spectrum information shown in Fig.~\ref{fig:E-R-PBC}, we conclude
that in the parameter regions with gapped Floquet spectra along the
imaginary quasienergy axis and $R=0$, the steady-state EE follows
an area-law scaling $S(L,L/2)\propto L^{0}$. Meanwhile, in the regions
with $R\in(0,1]$ and gapless Floquet spectra along the imaginary
axis, the steady-state EE follows a volume-law scaling $S(L,L/2)\propto L$.
Therefore, there should be entanglement transitions between volume-law
entangled and area-law entangled phases with the change of hopping
or gain/loss strength in our Floquet system. To confirm these entanglement
transitions, we present in Figs.~\ref{fig:EEdEE}(c) and \ref{fig:EEdEE}(d)
the gradient $g$ of steady-state EE, as obtained from the linear
fitting $S(L,L/2)\sim gL+s_{0}$, versus $J_{1}$ and $\gamma$. Multiple
area-law to volume-law (with $g=0\rightarrow>0$) and volume-law to
area-law (with $g>0\rightarrow=0$) entanglement phase transitions
are now clearly observable. Two notable features deserve to be further
emphasized.

First, with the increase of quenched intracell hopping amplitude $J_{1}$
from zero, we find a series of alternated transitions between volume-law
entangled and area-law entangled phases {[}Fig.~\ref{fig:EEdEE}(c){]}.
Similar patterns of alternated entanglement transitions can be obtained
with the variation of intercell hopping amplitude $J_{2}$ when $J_{1}$
is fixed. Therefore, we could induce and even engineer entanglement
phase transitions with high flexibility in non-Hermitian Floquet systems
by tuning a single control parameter, which are hardly achievable
in non-driven situations. The underlying physical picture is as follows.
Since the real part of quasienergy is a phase factor and defined modulus
$2\pi$, two quasienergy bands of a non-Hermitian Floquet system could
meet with each other and separate again at both
$E=0$ (center of the first quasienergy BZ) and $E=\pi$ (edge of
the first quasienergy BZ). Moreover, due to the $2\pi$-periodicity
of $E$, the values of Floquet quasienergies $E\mod2\pi$ in general
could not change monotonically with the increase or decrease of a
single system parameter. The combination of these two mechanisms then
allows the Floquet bands of our system to touch and re-separate along
the ${\rm Im}E$ axis sequentially at ${\rm Re}E=0$ and ${\rm Re}E=\pi$.
The final results are alternated entanglement phase transitions triggered
by a single driving parameter, as observed in Fig.~\ref{fig:EEdEE}(c).

Second, with the increase of gain/loss amplitude $\gamma$, the system
could first undergo a volume-law to area-law entanglement transition,
which is followed by reentering a volume-law entangled phase through
another entanglement transition at a larger $\gamma$, and finally
going back to a area-law entangled phase with the further raise of
$\gamma$ {[}Fig.~\ref{fig:EEdEE}(d){]}. Here, the non-Hermiticity-induced
reentrant transition from area-law entangled to volume-law entangled
phases is abnormal and usually not available in static non-Hermitian
systems. As the real part of quasienergy also depends on $\gamma$,
the reentrant entanglement transition observed here is due to the
presence of two possible gap-closing points at $E=0$ and $\pi$, the $2\pi$
periodicity of ${\rm Re}E$ and the non-monotonous dependence of Floquet
spectrum compositions on $\gamma$. Assisted by Floquet drivings,
the re-emerged volume-law entangled phase at large dissipation rates
may provide us with further room for protecting quantum information
against decoherence.

\begin{figure}
	\begin{centering}
		\includegraphics[scale=0.236]{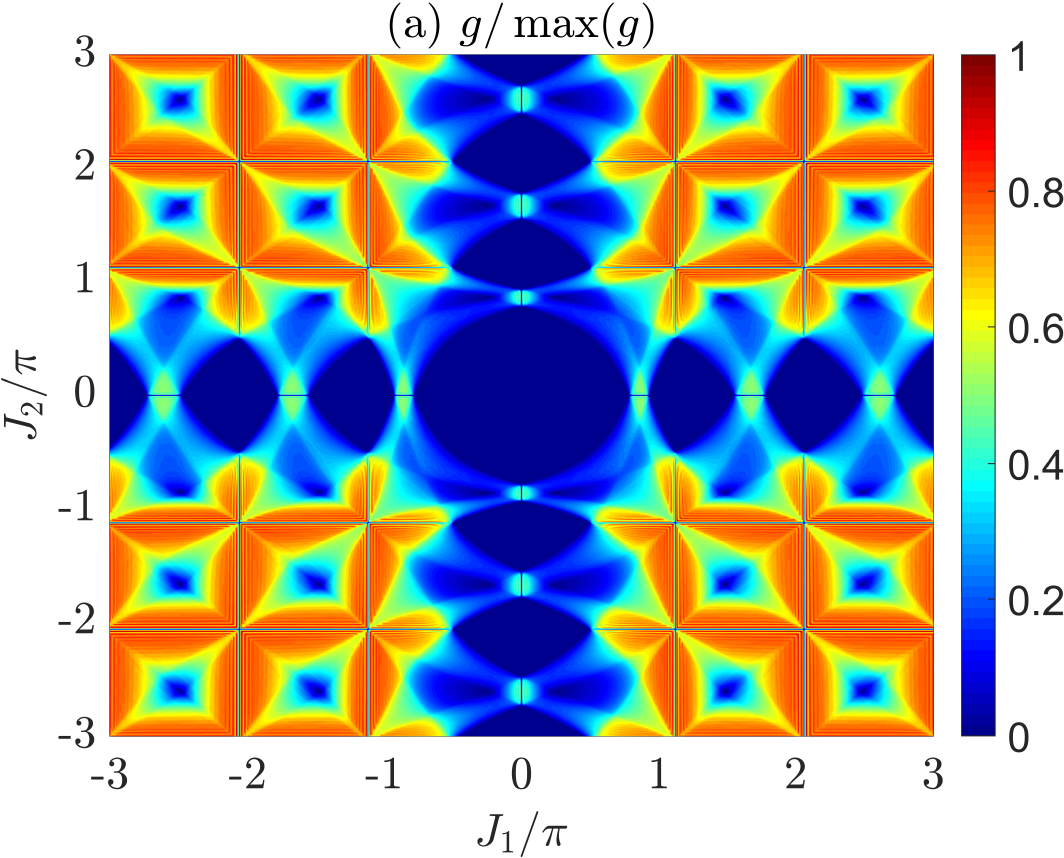}$\,\,$\includegraphics[scale=0.236]{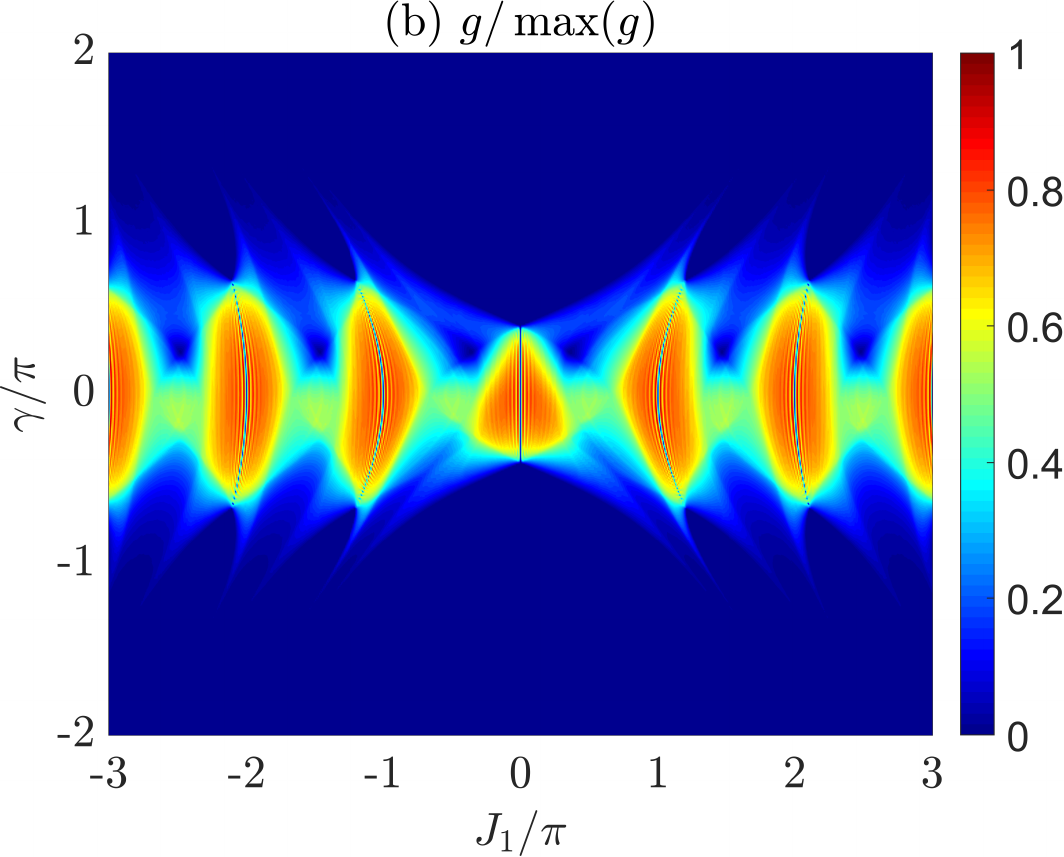}
		\par\end{centering}
	\caption{Entanglement phase diagrams vs $(J_{1},J_{2})$ and $(J_{1},\gamma)$.
		Other system parameters are $\gamma=0.5\pi$ for (a) and $J_{2}=2\pi/3$
		for (b). The gradient $g$ is obtained from the linear-fitting $S(L,l)\sim gL+s_{0}$
		of the steady-state EE with $l=L/2$ at half-filling and under PBC.
		The $\max(g)$ denotes the maximum of $g$ over the considered parameter
		space $(J_{1},J_{2})\in(-3\pi,3\pi)\times(-3\pi,3\pi)$ {[}$(J_{1},\gamma)\in(-3\pi,3\pi)\times(-2\pi,2\pi)${]}
		in (a) {[}(b){]}. \label{fig:EPTPhsDiag}}
\end{figure}
For completeness, we present in Fig.~\ref{fig:EPTPhsDiag} the gradient
$g$ extracted from the linear fitting $S(L,L/2)\sim gL+s_{0}$ of
the steady-state EE versus $L$ at different system parameters, which
constitutes the entanglement phase diagram of our periodically quenched
NHSSH model. A comparison between Figs.~\ref{fig:PTPhsDiag} and \ref{fig:EPTPhsDiag}
yield a nice consistency, i.e., the regions with $R=0$ (fully complex
Floquet spectra with finite dissipation gaps along ${\rm Im}E$) and
$R>0$ (partially complex or real Floquet spectra with no dissipation
gaps) are associated with area-law entangled and volume-law entangled
phases, respectively. Moreover, rich patterns of entanglement phase
transitions are observable over broad regions in the hopping and gain/loss
parameter spaces. Therefore, we conclude that the interplay between
periodic driving and non-Hermitian effects could not only generate
rich phases with different entanglement nature, but also trigger alternated
and reentrant entanglement transitions in examplary non-Hermitian
Floquet systems.

In Appendix \ref{sec:AppB}, we briefly discussed a model following a different driving protocol, in which the intra- and inter-cell hopping terms are both present within each half of the driving period. Alternated and reentrant entanglement transitions are also found for the system considered there, whose features are qualitatively the same as those reported in our main text.

\section{Conclusion and discussion}\label{sec:Sum}
In this work, we applied the idea of Floquet engineering to generate
and control entanglement phase transitions in non-Hermitian systems.
By applying time-periodic quenches to the hopping amplitudes of fermions
in a prototypical SSH model with balanced gain and loss, we found
alternated and reentrant entanglement transitions due to the combined
efforts of Floquet driving and non-Hermitian effects. System-size
scaling behaviors of steady-state EE were systematically analyzed
and entanglement phase diagrams were formulated for our considered
model. The alternated transitions between volume-law and area-law
entangled phases are due to driving-induced consecutive closings and
re-openings of Floquet dissipation gaps along the imaginary quasienergy
axis. The driving field also allows the composition of Floquet spectrum
(real vs complex) to change non-monotonically with the increase of
gain and loss strengths, yielding abnormal area-law to volume-law
reentrant transitions in the steady-state EE following the raise of non-Hermitian
effects. The alternated and reentrant entanglement transitions found
here are expected to be generic and observable in other driven non-Hermitian
systems. Our work thus unveiled the diversity and richness of entanglement
phase transitions in non-Hermitian Floquet systems. It further provided
a flexible route to induce and control entanglement phase transitions
in open systems via periodic driving fields.

Some remarks on the construction of our model and its experimental relevance are in order. First, in the early work of measurement-induced entanglement transitions \cite{EPT01,EPT02,EPT03}, a hybrid quantum circuit with a brick-wall structure in spacetime was considered as a prototypical setup. That system consists of a one-dimensional (1D) spin-$1/2$ (or qubit) chain with nearest-neighbor gates. Every discrete time period of the circuit has two layers. Each layer has $L/2$ gates, operating on the odd links in the first layer and the even links in the second layer (see the Fig.~1 of Ref.~\cite{EPT01}). From the perspective of Floquet system, the model considered in Ref.~\cite{EPT01} may be viewed as a spin-$1/2$ chain whose nearest-neighbor spin-spin couplings are switched on and off alternatively along odd and even bonds within each driving period. A comparison between that hybrid quantum circuit and the model introduced in our Sec.~\ref{sec:Mod} suggests a clear similarity regarding the employed driving protocols. The main differences are that in our case, each qubit is replaced by a lattice site, the coupling bewteen adjacent sites become staggered single-particle hopping, and the non-Hermitian effect is formally introduced by local loss instead of the projective measurements in Ref.~\cite{EPT01}. Therefore, upon suitable modifications, an experimental setup that could realize the circuit models in previous studies \cite{EPT01,EPT02,EPT03} may also be used to realize our model and explore entanglement transitions therein. Besides, as discussed in Appendix \ref{sec:AppB}, alternated and reentrant entanglement transitions can appear in our setup even if both intracell and intercell hopping terms are existent in each half of the driving period. The construction of our model thus captures the essential feature of entanglement transitions in a class of non-Hermitian Floquet system with PT symmetry.

In current experiments, periodically quenched non-Hermitian lattices with internal degrees of freedom can be implemented in cold atoms and photonic systems. In cold atoms, the SSH model and its dynamical modulations have been realized through various different strategies \cite{CdAtmSSH01,CdAtmSSH02,CdAtmSSH03,CdAtmSSH04}. For example, in a momentum-space lattice of cold atoms, intracell and intercell hopping terms can be alternatively switched on and off within a quantum-walk scheme \cite{PQSSH01}. The periodic quenching protocol in our Sec.~\ref{sec:Mod} could thus be implemented. Furthermore, local loss terms could be introduced by coupling momentum states to an auxiliary momentum-state bath or a different internal state that can be moved away by resonant light \cite{NHCdAtm01,NHCdAtm02,NHCdAtm03,NHCdAtm04,CdAtmNHExp01,CdAtmNHExp02,CdAtmNHExp03}. These approaches have been utilized to realize PT symmetry breaking and nonreciprocal transport in cold atoms. Meanwhile, as discussed in Sec.~\ref{sec:EPT}, the dynamics of our system is not affected if we only have local loss terms. Putting together, both the driving protocol and non-Hermitian effects of our system should be realizable via non-Hermitian quantum walks of cold atoms in momentum-space lattices, and our model may thus be engineered in such a setup. In photonic systems, 1D arrays of periodically driven resonators have been implemented to detect non-Hermitian Floquet band structures \cite{NHPho06}. In parallel with cold-atom quantum walks, photonic quantum walks could also realize alternated switching on and off of intra- and inter-cell hoppings, together with non-Hermitian effects introduced by polarization-dependent photon loss \cite{NHPho01,NHPho02,NHPho03,NHPho04,NHPho05}. Photonic setups thus provide another promising candidate for the realization of our Floquet non-Hermitian SSH model and the exploration of its dynamics. Putting together, our system and its entanglement transitions may be experimentally explorable in near-term quantum simulators.

In future work, it would be interesting to consider entanglement phase
transitions in non-Hermitian Floquet systems beyond one spatial dimension \cite{Ageev2021},
with impurities or disorder, under other driving protocols, and subject
to many-body interactions. Systematic analyses regarding the critical
behaviors of steady-state EE at entanglement transition points are
highly desired in non-Hermitian Floquet systems. The experimental
realization of our quenched non-Hermitian lattice and the detection
of entanglement phase transitions therein also constitute interesting
directions of future research.

\begin{acknowledgments}
	This work is supported by the National Natural Science Foundation of China (Grants No.~12275260, No.~12047503 and No.~11905211), the Fundamental Research Funds for the Central Universities (Grant No.~202364008), and the Young Talents Project of Ocean University of China.
\end{acknowledgments}

\appendix

\section{Stroboscopic EE and its numerical calculation}\label{sec:AppA}

Here we outline an approach that can be used to obtain the stroboscopic
EE for our system, which follows the method introduced in Ref.~\cite{EPTNH04}.
Compared with the algorithm outlined in the Appendix B of Ref.~\cite{EPTNH04}, our difference is just to choose the evolution time interval $\Delta t$ there as our driving period. This adjustment allows us to find the evolved state only stroboscopically.

In our study, the EE is obtained from a single wave function of noninteracting
particles. Following Eqs.~(\ref{eq:Psit}) and (\ref{eq:Psi0}), this state
vector of our system is given by $|\Psi(t)\rangle=|\Psi(\ell T)\rangle$.
The density matrix of the system at any stroboscopic time $t=\ell T$
then reads
\begin{equation}
	\rho(t)=|\Psi(t)\rangle\langle\Psi(t)|.\label{eq:DM}
\end{equation}
Since $|\Psi(t)\rangle$ is a normalized right state, $\rho(t)$
satisfies the general properties of a pure-state density matrix, i.e.,
$\rho(t)=\rho^{\dagger}(t)$, $\rho(t)=\rho^{2}(t)$, and ${\rm Tr}[\rho(t)]=1$.
If we decompose our 1D system into two spatially connected
segments X and Y, the reduced density matrix of the subsystem X can
be obtained by tracing out all the degrees of freedom belonging to
the subsystem Y, i.e.,
\begin{equation}
	\rho_{{\rm X}}(t)={\rm Tr}_{{\rm Y}}[\rho(t)].\label{eq:RDM}
\end{equation}
The von Neumann bipartite EE $S(t)$ between the subsystems X and
Y is defined in terms of $\rho_{{\rm X}}(t)$ as
\begin{equation}
	S(t)=-{\rm Tr}\{\rho_{{\rm X}}(t)\ln[\rho_{{\rm X}}(t)]\}.\label{eq:StApp}
\end{equation}

For free lattice models, a generic connection has been established
between the bipartite EE and the spectrum of single-particle correlator \cite{PeschelRev2009}. It is irrespective of whether the state evolves
following a Hermitian or a non-Hermitian Hamiltonian, so long as we
only consider normalized right state vectors. In the lattice representation,
the matrix element of single-particle correlator for our system is
given by the Eq.~(\ref{eq:CM}). By diagonalizing this correlation
matrix in the space of subsystem X, we find the set of its eigenvalues
$\{\zeta_{j}(\ell T)|j=1,2,...,2\ell-1,2\ell\}$. Here $\ell$ is
the number of unit cells belonging to the subsystem X, which has $2\ell$
sites in total since there are two sublattices within each unit cells.
Using these eigenvalues and following Ref.~\cite{PeschelRev2009}, we
can obtain the bipartite EE in terms of Eq.~(\ref{eq:EEt}).
Note in passing that our definitions and calculations of EE do not
employ biorthogonal state vectors in different Hilbert spaces (see
\cite{NHEEBio01,NHEEBio02,NHEEBio03} for discussions of the biorthogonal approach), making the
recipe of Ref.~\cite{PeschelRev2009} to be applicable.

The remaining issue is to evaluate the wave function $|\Psi(t)\rangle$
of noninteracting particles and find out the correlation matrix element
in Eq.~(\ref{eq:CM}) efficiently. Starting with the normalized state $|\Psi(t)\rangle$ {[}Eq.~(\ref{eq:Psit}){]}
at the stroboscopic time $t=\ell T$, we obtain the state after one
more evolution period as
\begin{alignat}{1}
	& |\Psi(t+T)\rangle\propto\hat{U}|\Psi(t)\rangle\nonumber \\
	= & \prod_{n=1}^{N}\left[\sum_{m=1}^{L}\sum_{s=A,B}[e^{-iH_{{\rm eff}}}{\cal U}]_{msn}(t)\hat{c}_{m,s}^{\dagger}\right]|\emptyset\rangle.\label{eq:Psitt}
\end{alignat}
Here $N$ counts the total number of fermions. $L$ is the number
of unit cells in the lattice. $H_{{\rm eff}}$ is a $2L\times2L$
matrix of the Floquet effective Hamiltonian $\hat{H}_{{\rm eff}}\equiv i\ln\hat{U}$
in the lattice representation. The normalized state at $t+T$ can
be obtained by performing the QR-decomposition, i.e.,
\begin{equation}
	e^{-iH_{{\rm eff}}}{\cal U}={\cal Q}{\cal R}.\label{eq:QR}
\end{equation}
Here ${\cal Q}$ is a $2L\times N$ matrix satisfying ${\cal Q}^{\dagger}{\cal Q}=1$.
${\cal R}$ is an $N\times N$ upper triangular matrix. The $2L\times N$
matrix ${\cal U}$ is isometry and it also satisfies ${\cal U}^{\dagger}{\cal U}=1$.
At the time $t+T$, the matrix ${\cal U}(t+T)$ is then given by
\begin{equation}
	{\cal U}(t+T)={\cal Q}.\label{eq:Utt}
\end{equation}
Note in passing that the matrix ${\cal U}(t=0)$ accounts the initial
distribution of fermions in the lattice. For the state $|\Psi_{0}\rangle$
in Eq.~(\ref{eq:Psi0}), the ${\cal U}(t=0)$ takes the explicit form
\begin{equation}
	[{\cal U}(0)]_{j,j'}=\delta_{j,2j'},\qquad j,j'=1,...,N,\label{eq:U0}
\end{equation}
where we have $L=N$ in the half-filled case. Following this approach,
we can find the ${\cal U}(\ell T)$ at any stroboscopic time $t=\ell T$.
The matrix elements of single-particle correlator $C(\ell T)$ can
be obtained as
\begin{equation}
	C_{ms,m's'}(\ell T)=[{\cal U}(\ell T){\cal U}^{\dagger}(\ell T)]_{m's',ms}.\label{eq:CM2}
\end{equation}
The EE can be finally extracted from the spectrum of $C(\ell T)$
according to Eq.~(\ref{eq:EEt}) in the main text. 

The approach considered here is
efficient for studying the long-time stroboscopic dynamics of
EE. It is also applicable to both clean and disordered noninteracting
fermionic systems and under different boundary conditions.
As reported in the main text, our calculations of the Floquet bands and EE give rise to consistent and physically reasonable predictions about the spectrum and entanglement transitions in our system. These can be observed in our Fig.~\ref{fig:E-R-PBC} vs Fig.~\ref{fig:EEdEE} (see also Fig.~\ref{fig:PTPhsDiag} vs Fig.~\ref{fig:EPTPhsDiag}). Such observations offer crosschecks for the correctness and consistency of our dynamical methodology.

\section{Another driving protocol}\label{sec:AppB}
In the main text, we studied a model whose intracell and intercell
hopping terms are separately turned on in two different halves of
each driving period. This protocol allows us to demonstrate the alternated
and reentrant entanglement transitions in non-Hermitian Floquet systems
with transparent theoretical and numerical analyses. In this Appendix,
we showcase that the phenomena reported in the main text are not restricted
to the piecewise quenching protocol considered there.

\begin{figure}
	\begin{centering}
		\includegraphics[scale=0.49]{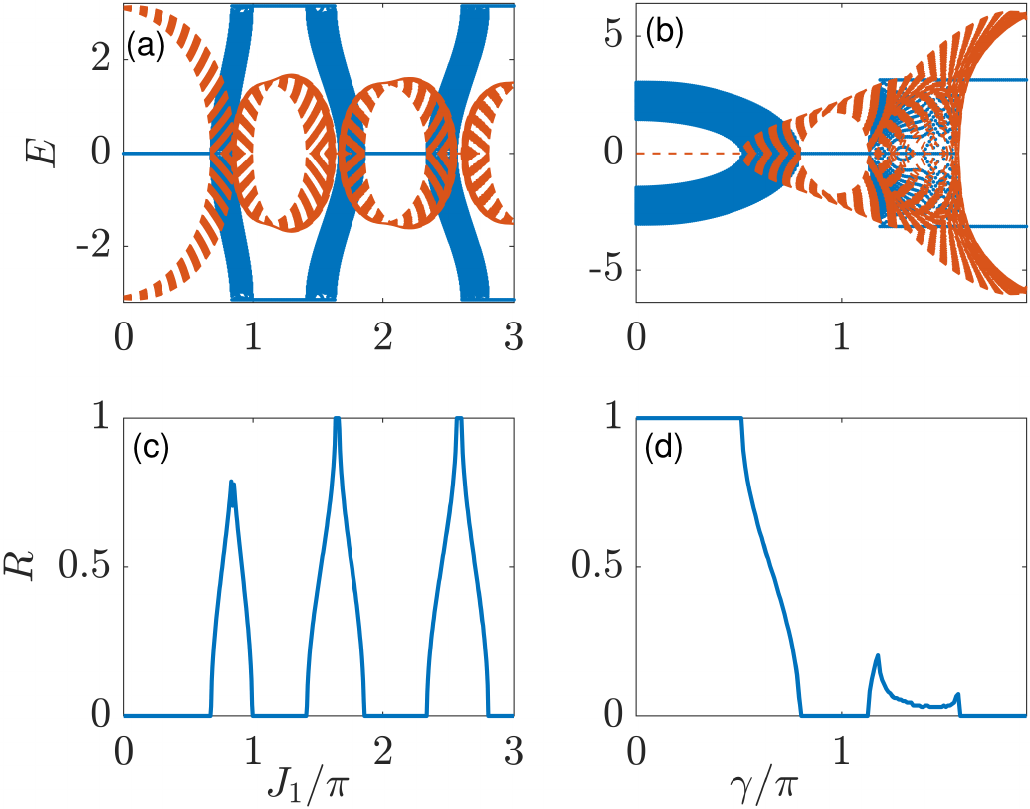}
		\par\end{centering}
	\caption{Floquet spectrum $E$ {[}(a), (b){]} and real-quasienergy ratios $R$
		{[}(c), (d){]} of $\hat{U}'$ {[}Eq.~(\ref{eq:Up}){]} vs the hopping
		amplitude $J_{1}$ and gain/loss parameter $\gamma$ under PBC. We
		set $J_{10}=J_{20}=\pi/20$ for all panels. Other system parameters
		are $(J_{2},\gamma)=(0.1\pi,0.5\pi)$ for (a), (c) and $(J_{1},J_{2})=(2.2\pi,2\pi/3)$
		for (b), (d). The solid points and dashed lines in (a) and (b) denote
		the real and imaginary parts of quasienergy. \label{fig:App1}}
\end{figure}

\begin{figure}
	\begin{centering}
		\includegraphics[scale=0.49]{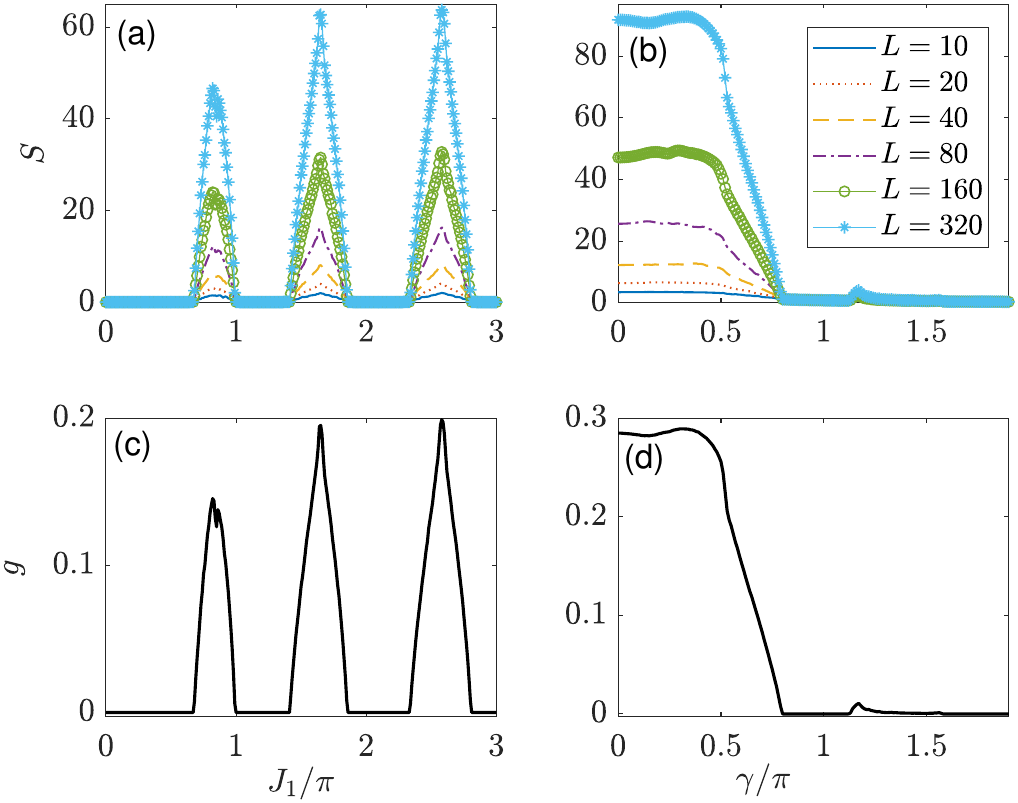}
		\par\end{centering}
	\caption{Reentrant entanglement transitions vs the hopping amplitude $J_{1}$
		{[}(a), (c){]} and gain/loss strength $\gamma$ {[}(b), (d){]} of $\hat{U}'$ {[}Eq.~(\ref{eq:Up}){]}. We
		set $J_{10}=J_{20}=\pi/20$ for all panels. Other system parameters
		are $(J_{2},\gamma)=(0.1\pi,0.5\pi)$ for (a), (c) and $(J_{1},J_{2})=(2.2\pi,2\pi/3)$
		for (b), (d). (a) and (b) show the steady-state EE $S(L,l)$, with
		$l=L/2$, vs $J_{1}$ and $\gamma$ for different lattice sizes $L$.
		(c) and (d) show the gradients $g$ extracted from the linear fitting
		$S(L,L/2)\sim gL+s_{0}$ of EE. \label{fig:App2}}
\end{figure}

As an example, we consider a situation in which both the intracell
and intercell hopping terms are present throughout each driving period.
The time-dependent Hamiltonian now takes the form
\begin{equation}
	\hat{H}'(t)=\begin{cases}
		\hat{H}'_{1} & t\in[\ell T,\ell T+T/2),\\
		\hat{H}'_{2} & t\in[\ell T+T/2,\ell T+T),
	\end{cases}\label{eq:Hpt}
\end{equation}
where $\hat{H}'_{1}=\hat{H}_{1}+\hat{H}_{10}$ and $\hat{H}'_{2}=\hat{H}_{2}+\hat{H}_{20}$.
$\hat{H}_{1}$ and $\hat{H}_{2}$ are Hamiltonians given by the
Eqs.~(\ref{eq:H1}) and (\ref{eq:H2}) in the main text. The additional terms read
\begin{equation}
	\hat{H}_{10}=J_{10}\sum_{n}(\hat{c}_{n,B}^{\dagger}\hat{c}_{n+1,A}+{\rm H.c.}),\label{eq:H10}
\end{equation}
\begin{equation}
	\hat{H}_{20}=J_{20}\sum_{n}(\hat{c}_{n,A}^{\dagger}\hat{c}_{n,B}+{\rm H.c.}).\label{eq:H20}
\end{equation}
The system now possesses intracell and intercell hopping with amplitudes
$J_{1}$ and $J_{10}$ ($J_{20}$ and $J_{2}$) in the first (second)
half of each driving period. The Floquet operator of this model is
given by
\begin{equation}
	\hat{U}'=e^{-i\hat{H}'_{2}}e^{-i\hat{H}'_{1}}.\label{eq:Up}
\end{equation}

Following the methods of Sec.~\ref{sec:Mod}, we obtain the spectra and real-quasienergy
ratios of the system described by the Floquet operator $\hat{U}'$
{[}Eq.~(\ref{eq:Up}){]}, as reported in Fig.~\ref{fig:App1}. We
find that with the variation of $J_{1}$, there are also alternated
transitions between real and complex Floquet spectra. Moreover, reentrant
transitions from complex to partially-real Floquet spectra are observed
with the increase of $\gamma$. The spectrum features of our Floquet
model considered in the main text can thus be qualitatively reproduced
after adding the intercell and intracell hopping terms contained in $\hat{H}_{10}$
and $\hat{H}_{20}$.

Referring to the approaches outlined in Sec.~\ref{sec:EPT} and Appendix \ref{sec:AppA}, we
obtain the steady-state EE $S(L,L/2)$ from the dynamics guided by
$\hat{U}'$ {[}Eq.~(\ref{eq:Up}){]} for different $L$, and further
extracting its scaling law vs $L$ at different hopping and non-Hermitian
parameters $J_{1}$ and $\gamma$ {[}Eq.~(\ref{eq:Up}){]}. As shown
in Fig.~\ref{fig:App2}, alternated and reentrant entanglement transitions
are clearly observed, and their features are qualitatively identical
to those presented in the Fig.~\ref{fig:EEdEE} for our original model $\hat{U}$.

Therefore, we conclude that the rich and unique features of spectrum
and entanglement transitions in our non-Hermitian Floquet system could
emerge under different quenching protocols (here with either $J_{10}=J_{20}=0$
or $J_{10},J_{20}\neq0$). It remains an interesting issue to explore
other types of Floquet entanglement transitions under different classes
of periodic driving fields (e.g., harmonic, delta-kicking) in future
studies.

\end{document}